\documentclass[aps,twocolumn,pra,superscriptaddress,amsmath,amssymb]{revtex4-2} 
\usepackage{dcolumn}
\usepackage{bm}
\usepackage{amsmath}
\usepackage{mathrsfs}
\usepackage{txfonts}
\usepackage[T1]{fontenc}
\usepackage{xspace}
\usepackage{ulem}
\usepackage{comment}
\usepackage{braket}
\usepackage{lipsum}
\usepackage{mathtools}
\newcommand{\means}[1]{\langle#1\rangle}

\newcommand{\pdiff}[2]{\frac{\partial {#1}}{\partial {#2}}}

\ifx\pdfoutput\undefined
\usepackage[dvipdfmx]{graphicx}
\usepackage[dvipdfmx]{hyperref}
\usepackage[dvipdfmx]{color}
\usepackage[dvipdfmx]{xcolor}
\else
\usepackage{graphicx}
\usepackage{hyperref}
\usepackage{color}
\usepackage{xcolor}
\usepackage[version=3]{mhchem}
\fi
\hypersetup{
        colorlinks=true,
        citecolor=blue,
        urlcolor=blue,
        linkcolor=blue
}

\begin{document}
\let\emph\textit

\title{
Thermal Hall effect incorporating magnon damping in localized spin systems
}
\author{Shinnosuke Koyama}
\author{Joji Nasu}
\affiliation{
  Department of Physics, Tohoku University, Sendai, Miyagi 980-8578, Japan
}

\date{\today}
\begin{abstract}
We propose a theory for thermal Hall transport mediated by magnons to address the impact of their damping resulting from magnon-magnon interactions in insulating magnets.
This phenomenon is anticipated to be particularly significant in systems characterized by strong quantum fluctuations, exemplified by spin-1/2 systems. 
Employing a nonlinear flavor-wave theory, we analyze a general model for localized electron systems and develop a formulation for thermal conductivity based on a perturbation theory, utilizing bosonic Green's functions with a nonzero self-energy.
We derive the expression of the thermal Hall conductivity incorporating magnon damping.
To demonstrate the applicability of the obtained representation, we adopt it to two $S=1/2$ quantum spin models on a honeycomb lattice.
In calculations for these systems, we make use of the self-consistent imaginary Dyson equation approach at finite temperatures for evaluating the magnon damping rate.
In both systems, the thermal Hall conductivity is diminished due to the introduction of magnon damping over a wide temperature range.
This effect arises due to the smearing of magnon spectra with nonzero Berry curvatures.
We also discuss the relation to the damping of chiral edge modes of magnons.
Our formulation can be applied to various localized electron systems as we begin with a general Hamiltonian for these systems.
Our findings shed light on a new aspect of topological magnonics emergent from many-body effects and will stimulate further investigations on the impact of magnon damping on topological phenomena.
\end{abstract}
\maketitle


\section{Introduction}
\label{introduction}

In condensed matter physics, the concept of topology provides profound insights into exotic electronic structures and the phenomena that arise from them.
The topology of bands, formed by Bloch electrons, is characterized by a topological invariant defined for each band.
The presence of a non-zero topological invariant predicts the emergence of gapless edge modes and the quantization of Hall conductivity in insulating states~\cite{thouless1982,kohmoto1985}.
This concept has been extended to include systems composed not only of fermionic particles, such as electrons, but also those with bosonic quasiparticles like phonons~\cite{sheng2006,zhang2010,zhang2011,qin2012,saito2019,wang2020,chen2021,ding2022} and magnons~\cite{katsura2010,matsumoto2011,mcclarty2022,zhang2023_arxiv,shindou2013,mook2014,matsumoto2014,li2016,zyuzin2016_prl}.
In bosonic systems, it is also possible to introduce the Berry curvature and topological invariants, such as the Chern number, for each band, similar to fermionic systems. Because bosonic quasiparticles are charge-neutral, the Hall effect does not manifest in these systems.
Instead, these quasiparticles can carry heat, implying the potential for the emergence of a thermal Hall effect in systems with topologically nontrivial band structures~\cite{sheng2006,zhang2010,zhang2011,qin2012,katsura2010,matsumoto2011,mcclarty2022,zhang2023_arxiv}.
For instance, in localized electron systems with spin degrees of freedom, it has been proposed that anisotropic spin interactions, such as Dzyaloshinskii-Moriya (DM) interactions~\cite{moriya1960} and Kitaev couplings~\cite{kitaev}, can induce a nonzero Chern number in magnon bands
~\cite{zyuzin2016_prb,zyuzin2016_prl,mcclarty2018,joshi2018,laurell2018,malki2019,li2020_prr,koyama2021,chuan_zhang2021_prb,chern2021_prl,chern2020_prr,zhang2021_prb,zhang2023,mook2021,mook2014,chen2023-prb,mook2023}.
Such magnons, termed topological magnons, have recently garnered significant attention~\cite{mcclarty2022,zhang2023_arxiv}.
In fact, the thermal Hall effect originating from magnons has been observed in materials with pyrochlore~\cite{onose2010,ideue2012,hirschberger2015_science}, honeycomb~\cite{czajka2023,zhang2021_prl}, and kagome structures~\cite{hirschberger2015_prl,Akazawa2020}.

Thus far, the topological properties of magnons, elementary excitations arising from a magnetic order in localized spin systems, have been conducted within the framework of linear spin-wave theory.
This approximation enables the representation of the thermal Hall coefficient through the Berry curvature of magnon bands, incorporating a contribution beyond the conventional Kubo formula.
This contribution is known as the heat magnetization arising from the orbital motion of magnons~\cite{matsumoto2011,shindou2013,matsumoto2014,murakami2017}.
The formalism based on the free-magnon picture has successfully explained experimental results, such as the magnetic-field dependence of a thermal Hall coefficient in $\mathrm{Lu_2 V_2 O_7}$~\cite{onose2010}.
Furthermore, this approach has been expanded to include calculations of other topological phenomena, including the spin Nernst effect and nonlinear responses~\cite{zyuzin2016_prb,zyuzin2016_prl,li2020_prr,mook2019,kondo2022,wang2022}. 
However, the spin-wave theory beyond the linear approximation indicates the presence of magnon-magnon interactions due to quantum fluctuations~\cite{holstein1940}.
These interactions become notably significant in systems with short spin lengths and when a large number of magnons are thermally excited. 
Such magnon-magnon interactions could play a crucial role in topological thermal transport phenomena, as magnons occupy bands with finite Berry curvature only at finite temperatures.
Furthermore, the fact that magnons are bosons emphasizes the significance of their interactions.
Quasiparticles that describe elementary excitations as bosons do not obey the conservation law of particle numbers, owing to their zero chemical potential. 
This leads to magnon-magnon interactions that do not conserve particle numbers, resulting in the decay of high-energy magnons even at low temperatures~\cite{harris1971,zhitomirsky1999,costa2000,chernyshev2006,chernyshev2009,mourigal2010,mourigal2013,zhitomirsky2013,maksimov2016_prb,maksimov2016_prl,chernyshev2016,winter2017_nc,mcclarty2018,mcclarty2019,kim2019,smit2020,mook2020,mook2021,maksimov2020,maksimov2022,bai2023,koyama2023,habel2024,koyama2023_NPSM}.
It has been suggested that this effect also influences the topological properties of magnons
~\cite{chernyshev2016,mcclarty2018,mcclarty2019,mook2020,mook2021,chen2023-prb,heinsdorf2023_arxiv}, particularly for the damping of the chiral edge modes~\cite{koyama2023,habel2024}, potentially suppressing the thermal Hall effect.
Therefore, to evaluate the thermal Hall conductivity, it is essential to consider the effects of magnon-magnon interactions properly.
This contrasts with that in electronic systems at low temperatures, where chiral edge modes remain robust against such interactions~\cite{hasan2010}.

Recent experimental results suggest that such magnon-magnon interactions significantly influence the thermal Hall effect. 
For instance, it has been reported that the measured values of thermal Hall conductivity in the layered material $\mathrm{Cr_2Ge_2Te_6}$ with a honeycomb structure exhibiting ferromagnetic order are considerably lower than those obtained by theoretical calculations under the linear spin-wave theory~\cite{choi2023}.
Furthermore, in the Shastry-Sutherland model, the elementary excitations termed triplons, which are bosonic quasiparticles similar to magnons, have been theoretically predicted to contribute to the thermal Hall conductivity within the free-particle approximation~\cite{romhanyi2015,mcclarty2017topological}.
However, experiments have not detected a thermal Hall effect in the candidate material $\mathrm{SrCu_2 (BO_3)_2}$~\cite{suetsugu2022}.
These studies imply that the discrepancies between theory and experiment may be attributed to neglecting interaction effects between magnons in the theoretical calculations.
Nevertheless, formulating a framework for the thermal Hall effect extending beyond the free-magnon approximation remains challenging, partly due to the complexity of considering the contribution of heat magnetization from magnons to the thermal Hall conductivity, in addition to the calculations from the Kubo formula.

In this paper, we formulate the thermal Hall conductivity in the presence of magnon damping in localized electron systems to elucidate the effect of magnon-magnon interactions on the thermal Hall effect.
Beginning with a general Hamiltonian for localized electron models, we adopt a mean-field (MF) approximation assuming a long-range order and introduce magnons as elementary excitations through the Holstein-Primakoff transformation.
This transformation not only produces a bilinear term of bosonic operators but also brings about additional terms responsible for magnon-magnon scattering.
We introduce the magnon Green's function, treating the former as an unperturbed term and the latter as a perturbation term.
We consider the effect of magnon damping as the imaginary part of the self-energy.
By assuming this part to be nonzero, we derive the expression for the thermal Hall conductivity incorporating magnon damping.
We apply this framework to the Kitaev model under a magnetic field and an $S=1/2$ spin model with Heisenberg and DM interactions, calculating the temperature dependence of thermal Hall conductivity in the presence of magnon damping.
The results reveal that the value of thermal Hall conductivity is significantly suppressed when magnons in bands with large Berry curvature decay strongly.
Our findings suggest that magnon damping plays a crucial role in thermal Hall conductivity in a wide temperature range.

This paper is organized as follows.
In Sec.~\ref{boson hamiltonian}, we introduce the model Hamiltonian and review the calculation method.
The MF approximation and Holstein-Primakoff transformation are presented in Secs.~\ref{method:MF} and \ref{method:GHP}, respectively, to introduce magnons as bosonic quasiparticules.
Based on the Holstein-Primakoff transformation, the spin-wave Hamiltonian is obtained in Sec.~\ref{method:LSW}.
In Sec.~\ref{sec:Green's}, we show the details of the perturbation theory based on the bosonic Green's function for magnons.
Analytical properties of the Green's function are also discussed in this section.
In Sec.~\ref{sec:thermal-transport}, we formulate the thermal conductivity using the bosonic Green's function.
We briefly present the general theory of thermal transport in Sec.~\ref{sec:intro-thermal-conductivity}.
The representation of thermal conductivity is shown in Sec.~\ref{sec:Sxy}.
In Sec.~\ref{sec:appro-green}, we present the expression of thermal Hall conductivity where the imaginary part of the self-energy is taken into account.
Section~\ref{sec:thermal-Hall-damping} shows the fundamental properties of thermal Hall conductivity in the presence of the magnon damping.
In Sec.~\ref{sec:result}, we show the calculation results of the thermal Hall conductivity using our framework in the following two quantum spin models: the Kitaev model under a magnetic field (Sec.~\ref{sec:Kitaev}) and an $S=1/2$ spin model with Heisenberg and DM interactions (Sec.~\ref{sec:Heisenberg-DM}).
Finally, Sec.~\ref{summary} is devoted to summary and discussion.

\section{Model and Method}
\label{boson hamiltonian}

\subsection{Mean-field theory}
\label{method:MF}

In this section, we briefly review the MF approximation.
We start from a general localized electron model, which is represented by
\begin{align}
  \label{eq:H}
  \mathcal{H}= \frac{1}{2}\sum_{i,j} \sum_{\alpha \beta}
  {J}^{\alpha\beta}_{ij} \mathcal{O}_{i}^{\alpha} \mathcal{O}_{j}^{\beta} 
  - \sum_{i}\sum_{\alpha} h_{i}^{\alpha}\mathcal{O}^{\alpha}_{i},
\end{align}
where $\mathcal{O}_i^\alpha$ is the $\alpha$ component of the local operator defined at site $i$, and $J_{ij}^{\alpha\beta}$ represents the exchange matrix between the operators $\mathcal{O}_{i}^\alpha$ and $\mathcal{O}_{j}^\beta$.
We consider $\mathscr{N}$ local states for each site.
The last term of Eq.~\eqref{eq:H} is the one-body term with the local field $h_{i}^\alpha$.
In the MF theory, Eq.~\eqref{eq:H} is divided to
\begin{align}
    \mathcal{H} = \mathcal{H}^{\mathrm{MF}} + \mathcal{H}',
\end{align}
where the first term represents the MF Hamiltonian, which is given by
\begin{align}
  \mathcal{H}^{\mathrm{MF}} = \sum_{i} \mathcal{H}_{i}^{\text{MF}} + \text{const.}
\end{align}
The local MF Hamiltonian $\mathcal{H}^{\mathrm{MF}}_{i}$ at site $i$ is represented as
\begin{align}
  \mathcal{H}_{i}^{\mathrm{MF}} = \sum_{\alpha} \left( \sum_{l'}^{M}\sum_{j\in l'}^{N_{u}/M} \sum_{\beta} 
  J^{\alpha\beta}_{ij} 
  \langle \mathcal{O}^{\beta} \rangle_{l'} 
  - h_{i}^{\alpha} \right) \mathcal{O}^{\alpha}_{i},
  \label{eq:lcoal-mf}
\end{align}
where $N_{u}$ and $M$ are the number of unit cells and sublattices, respectively.
A sublattice here refers to the set of sites where the same MF is assumed.
The expectation value $\langle \mathcal{O}^\alpha \rangle_{l} = \braket{0;i|\mathcal{O}^\alpha|0;i}$ is introduced for the ground state $\ket{0;i}$ 
of the local Hamiltonian $\mathcal{H}^{\text{MF}}_{i}$ with site $i$ belonging to sublattice $l$.
This ground state is obtained by diagonalizing $\mathcal{H}^{\mathrm{MF}}_{i}$.
Moreover, we obtain the $m$-th excited states $\ket{m;i}$ for $m=1,2,\cdots , \mathscr{N} - 1$ of $\mathcal{H}^{\mathrm{MF}}_{i}$ in a similar manner.

\subsection{Generalized Holstein-Primakoff transformation}
\label{method:GHP}

In this section, we rewrite the original Hamiltonian using bosons to describe the elementary excitations from the MF ground state.
We expand the local operator using the eigenstates of the local MF Hamiltonian at site $i$ in sublattice $l$ as
\begin{align}
  \label{eq:delO_2}
  \mathcal{O}^\alpha_{i} \ = \ \sum_{m,m' =0}^{\mathscr{N} -1}
  X^{mm'}_{i} \mathcal{O}_{mm';l}^{\alpha},
\end{align}
where $X^{mm'}_{i}\equiv\ket{m;i}\bra{m';i}$ and $\mathcal{O}_{mm';l}^{\alpha}=\braket{m;i|\mathcal{O}^{\alpha}|m';i}$, which depends only on the sublattice index $l$ to which site $i$ belongs.
$X_{i}^{mm'}$ is represented by bosons using the generalized Holstein-Primakoff transformation, which is known as a flavor-wave theory~\cite{joshi1999, kusunose2001,Lauchli2006,Tsunetsugu2006,Kim_flavor-wave2017, nasu2021, koyama2021,koyama2023}.
We introduce $\mathscr{N} - 1$ bosonic operators $a_{mi}^\dagger$ ($a_{mi}$) with $m=1,2,\cdots , \mathscr{N} - 1$ for each site.
For $m\geq 1$, $X^{0m}_{i}$ and $X^{m0}_{i}$ are given by
\begin{align}\label{eq:Xm0-operator}
  X_{i}^{m0} = a_{mi}^\dagger
  \left(
    \mathcal{S} - \sum_{n=1}^{\mathscr{N} - 1} a_{ni}^\dagger a_{ni}^{}
  \right)^{1/2}, \quad
  X_{i}^{0m} = \left( X_i^{m0} \right)^\dagger.
\end{align}
Here, ${\cal S}$ is introduced as 
\begin{align}\label{eq:sum-unity}
  \mathcal{S}= X_i^{00}+ \sum_{n=1}^{\mathscr{N} - 1} a_{ni}^\dagger a_{ni}^{},
\end{align}
and it should be unity because of $\sum_{m=0}^{\mathscr{N} -1}X_i^{mm}=1$.
For $1\leq m,m'$, $X_i^{mm'}$ is given by
\begin{align}
  \label{eq:excited-excited}
  X_i^{mm'} = a_{mi}^\dagger a_{m' i}^{}.
\end{align}
When the number of bosons is small enough, one can expand the square root in Eq.~\eqref{eq:Xm0-operator}  with respect to $1/\mathcal{S}$~\cite{joshi1999,kusunose2001,nasu2021,koyama2021,nasu2022,koyama2023}.
Using the expression, $\mathcal{H}$ is represented by the bosons and expanded for $1/{\cal S}$ as
\begin{align}
  \label{eq:bosonic_H}
  \mathcal{H} =
    \mathcal{S}\left(\mathcal{H}_{0}+  \frac{1}{\sqrt{\mathcal{S}}}\mathcal{H}_3 + \frac{1}{\mathcal{S}}\mathcal{H}_4 + O(\mathcal{S}^{-3/2})
  \right) + \text{const.},
\end{align}
where $\mathcal{H}_{0}$ is the bilinear term consisting of bosonic operators and $\mathcal{H}_{3}$ and $ \mathcal{H}_{4}$ are the terms composed of the three and four bosonic operators.
The explicit expression for $\mathcal{H}_{0}$ is shown the next section.
Note that, while $a_{mi}^\dagger$ and $a_{mi}$ do not appear alone because of the stable condition of the MF solution, other odd-order terms are allowed to appear in the Hamiltonian~\cite{zhitomirsky2013}.

\subsection{Flavor-wave theory}
\label{method:LSW}

In the bosonic representation in Eq.~\eqref{eq:bosonic_H}, $\mathcal{H}_{0}$ is regarded as a noninteracting Hamiltonian.
This is written as
\begin{align}
\mathcal{H}_{0} = &\sum_{l}^{M} \sum_{u}^{N_u} \sum_{m=1}^{\mathscr{N}-1} \Delta E_{m}^{l} a_{m(l,u)}^\dagger a_{m(l,u)}^{}\nonumber\\
     &+
\sum_{ll'}^{M}\sum_{uu'}^{N_u}\sum_{\alpha\beta}\sum_{mm'=1}^{\mathscr{N}-1}
    \frac{J_{ij}^{\alpha\beta}}{2}\nonumber\\
    &\quad\times\left( \mathcal{O}_{m0;l}^{\alpha} a_{m(l,u)}^\dagger + \mathrm{H.c.} \right) \left( \mathcal{O}_{m'0;l'}^{\beta} a_{m'(l',u')}^\dagger + \mathrm{H.c.} \right),
    \label{eq:H0-messy}
\end{align}
where $\Delta E_{m}^{l}$ is the energy difference between the excited state and ground state of the local MF Hamiltonian at site $i$ belonging to sublattice $l$.
The site label $i$ is expressed by the two indices $(l,u)$ with unit cell $u$ and sublattice $l$.
We also introduce $s=(l,m)$ as the composite index of sublattice $l$ and local excited state $m$ with $N=M(\mathscr{N}-1)$ being the number that $s$ can take. 
Note that $N$ is the number of branches for the collective modes~\cite{nasu2021}.
Then, $\mathcal{H}_{0}$ is represented as
\begin{align}
    \mathcal{H}_{0}
    =
    \frac{1}{2}\sum_{uu'ss'} 
    \Bigg[
    &\left(\mathcal{M}^{11}_{uu'}\right)_{ss'} a_{us}^\dagger a_{u's'}^{} + \left(\mathcal{M}_{uu'}^{12}\right)_{ss'} a_{us}^\dagger a_{u's'}^\dagger\nonumber\\
    &+ \left(\mathcal{M}^{21}_{uu'}\right)_{ss'} a_{us}^{} a_{u's'}^{} + \left(\mathcal{M}^{22}_{uu'}\right)_{ss'} a_{us}^{} a_{u's'}^\dagger
    \Bigg],
    \label{eq:H0-expand}
\end{align}
where $\mathcal{M}_{uu'}^{11},\mathcal{M}_{uu'}^{12},\mathcal{M}_{uu'}^{21}$, and $\mathcal{M}_{uu'}^{22}$ are the $N\times N$ matrices satisfying the following relations~\cite{nasu2021}:
\begin{align}
    \mathcal{M}_{uu'}^{11} &= \left(\mathcal{M}_{u'u}^{11} \right)^{\dagger} = \left(\mathcal{M}_{u'u}^{22} \right)^{T} = \left(\mathcal{M}_{uu'}^{22} \right)^{*},\label{eq:M11}\\
    \mathcal{M}_{uu'}^{12} &= \left(\mathcal{M}_{u'u}^{12} \right)^{T} = \left(\mathcal{M}_{u'u}^{21} \right)^{\dagger} = \left(\mathcal{M}_{uu'}^{21} \right)^{*}.
    \label{eq:M12}
\end{align}
We also introduce the $2N$-dimensional vector $\mathcal{A}_{u}^\dagger$, which is given by
\begin{align}
  \mathcal{A}_{u}^{\dagger} = \left(a_{u,1}^{\dagger}\ a_{u,2}^{\dagger}\ \cdots 
  \ a_{u,N}^{\dagger} 
  \ a_{u,1}^{} \ a_{u,2}^{} \ \cdots \ a_{u,N}^{} \right).
\end{align}
Then, $\mathcal{H}_{0}$ is rewritten as follows:
\begin{align}
  \label{eq:Hamil-A-real}
  \mathcal{H}_{0} = \frac{1}{2} \sum_{uu'}^{N_{u}} \mathcal{A}_{u}^\dagger {\cal M}_{uu'}^{} \mathcal{A}_{u'}^{}.
\end{align}
We introduce the $2N\times 2N$ Hermitian matrix $\mathcal{M}_{uu'}$, which is given by
\begin{align}
    \label{eq:Mdelta}
    \mathcal{M}_{uu'} =
    \begin{pmatrix}
        \mathcal{M}_{uu'}^{11} & \mathcal{M}_{uu'}^{12}\\[8pt]
        \left(\mathcal{M}_{uu'}^{12}\right)^{*} & \left(\mathcal{M}_{u'u}^{11}\right)^{T}
    \end{pmatrix}.
\end{align}
Note that $\mathcal{M}_{uu'}$ depends only on the relative positions of unit cells $u$ and $u'$.
By introducing the Fourier transformation of $a_{u,s}$ with respect to $u$, the Hamiltonian $\mathcal{H}_{0}$ is formally written as
\begin{align}
  \label{eq:Hamil-A}
  \mathcal{H}_{0} = \frac{1}{2} \sum_{\bm{k}} \mathcal{A}_{\bm{k}}^\dagger {\cal M}_{\bm{k}}^{} \mathcal{A}_{\bm{k}}^{},
\end{align}
where ${\cal M}_{\bm{k}}$ is a $2N\times 2N$ Hermitian matrix.
The sum of $\bm{k}$ is taken in the first Brillouin zone.
The $2N$-dimensional vector $\mathcal{A}_{\bm{k}}^{\dagger}$ is given by
\begin{align}
  \mathcal{A}_{\bm{k}}^{\dagger} = \left(a_{\bm{k},1}^{\dagger}\ a_{\bm{k},2}^{\dagger}\ \cdots 
  \ a_{\bm{k},N}^{\dagger} 
  \ a_{-\bm{k},1}^{} \ a_{-\bm{k},2}^{} \ \cdots \ a_{-\bm{k},N}^{} \right),
\end{align}
where $a_{\bm{k},s}$
is the Fourier transformation of $a_{mi}$, 
which is represented by
\begin{align}
  a_{\bm{k},s}^{}= \sqrt{\dfrac{1}{N_u}}\sum_{u} a_{u,s}^{} e^{-i\bm{k}\cdot \bm{r}_i}.
\end{align}
Here, we replace the index $(mi)$ in $a_{mi}$ to $(u,s)$, and $\bm{r}_i$ is the position of site $i$ belonging to sublattice $l$ in unit cell $u$.
By introducing the representative position of unit cell $\tilde{\bm{r}}_u$,
$\bm{r}_i$ is represented as $\bm{r}_i = \tilde{\bm{r}}_u + \tilde{\bm{\delta}}_s$, where $\tilde{\bm{\delta}}_s$ for 
$s=1,2,\cdots,N$ is a relative vector from $\tilde{\bm{r}}_u$ to $\bm{r}_i$.
Note that $\tilde{\bm{\delta}}_{s}$ depends only on the sublattice index $l$ in $s=(l,m)$.
$\mathcal{M}_{\bm{k}}$ is given as 
\begin{align}
    \left(\mathcal{M}_{\bm{k}}\right)_{ss'} = \sum_{u} 
    \exp \left[-i\bm{k}\cdot (\tilde{\bm{r}}_{u} + \tilde{\bm{\delta}}_{s} - \tilde{\bm{r}}_{u'}-\tilde{\bm{\delta}}_{s'})\right]
    \left(\mathcal{M}_{uu'}\right)_{ss'}.
\end{align}

We diagonalize ${\cal M}_{\bm{k}}^{}$ by applying the Bogoliubov transformation as $\mathcal{E}_{\bm{k}}^{}=T_{\bm{k}}^\dagger \mathcal{M}_{\bm{k}}^{} T_{\bm{k}}^{}$,
where $T_{\bm{k}}$ is a paraunitary matrix, 
which sastisfies the relation $T_{\bm{k}}^{}\sigma_3 T_{\bm{k}}^\dagger = T_{\bm{k}}^{\dagger}\sigma_3 T_{\bm{k}}^{} = \sigma_3$
with the paraunit matrix
$
\sigma_3 \equiv
\begin{psmallmatrix}
\bm{1}_{N\times N} & 0\\
0 & -\bm{1}_{N\times N}
\end{psmallmatrix}
$
, where $\bm{1}_{N\times N}$ is the $N\times N$ unit matrix.
$\mathcal{E}_{\bm{k}}$ is the diagonal matrix given by
$\mathcal{E}_{\bm{k}} = \mathrm{diag}\{\varepsilon_{\bm{k},1}, \varepsilon_{\bm{k},2}, \cdots, \varepsilon_{\bm{k},N}, 
\varepsilon_{-\bm{k},1}, \varepsilon_{-\bm{k},2}, \cdots, \varepsilon_{-\bm{k},N}\}$~\cite{colpa}.
Using this transformation, we rewrite the Hamiltonian as the following diagonalized form:
\begin{align}
  \label{eq:Hamil-B}
  \mathcal{H}_{0}= \frac{1}{2}\sum_{\bm{k}} \mathcal{B}^{\dagger}_{\bm{k}} \mathcal{E}_{\bm{k}} \mathcal{B}_{\bm{k}}^{},
\end{align}
Here, we introduce the set of bosonic operators $\mathcal{B}_{\bm{k}}^{}  = T_{\bm{k}}^{-1} \mathcal{A}_{\bm{k}}^{}$, which is given by
\begin{align}
  \mathcal{B}_{\bm{k}}^{\dagger} =\left(b_{\bm{k},1}^{\dagger}\ b_{\bm{k},2}^{\dagger}\ \cdots 
  \ b_{\bm{k},N}^{\dagger} 
  \ b_{-\bm{k},1}^{} \ b_{-\bm{k},2}^{} \ \cdots \ b_{-\bm{k},N}^{} \right).
\label{eq:Bogoliubov-Bop}
\end{align}
Note that $\mathcal{B}_{\bm{k}}$ satisfies the following commutation relation:
\begin{align}
    \label{eq:commu-B}
    \left[ \mathcal{B}_{\bm{k},\eta}^{}, \mathcal{B}_{\bm{k},\eta'}^\dagger \right] = \sigma_{3,\eta} \delta_{\eta,\eta'}.
\end{align}
While $\mathcal{H}_0$ in Eq.~\eqref{eq:Hamil-B} is written as a free-boson Hamiltonian, higher order terms such as $\mathcal{H}_{3}$ and $\mathcal{H}_{4}$ in Eq.~\eqref{eq:bosonic_H} describe interactions between bosons.

\subsection{Perturbation theory using Green's functions}

\label{sec:Green's}

In this section, we introduce the method addressing higher-order terms describing the interactions between bosons introduced in Sec.~\ref{method:GHP}.
Here, the bosonic representation of the Hamiltonian in Eq.~\eqref{eq:bosonic_H} is split into two terms: ${\cal H}/\mathcal{S}=\mathcal{H}_{0} + \mathcal{H}_{\rm int}$, where $\mathcal{H}_{\rm int}$ is the interactions between bosons, as shown in Sec.~\ref{method:GHP}.
The higher-order contributions $\mathcal{H}_{\rm int}$ are incorporated by using the perturbation theory, where $\mathcal{H}_{0}$ is regarded as an unperturbed term~\cite{zhitomirsky1999,mourigal2010,chernyshev2009}.
The perturbation term is given by
$
\mathcal{H}_{\text{int}} = \mathcal{H}_3/\sqrt{\mathcal{S}} + \mathcal{H}_4/\mathcal{S} + O(\mathcal{S}^{-3/2})
$.

To perform the perturbation expansion systematically, we employ the Green's function approach~\cite{mahan}.
We define the temperature Green's function for the Bogoliubov bosons $\mathcal{B}_{\bm{k},\eta}$ as
\begin{align}
  \mathcal{G}_{\bm{k},\eta\eta'}(\tau) = -\big\langle T_{\tau} \mathcal{B}_{\bm{k},\eta}^{}(\tau) \mathcal{B}_{\bm{k},\eta'}^\dagger \big\rangle,\label{eq:definition-Green}
\end{align}
where $T_{\tau}$ is the time-ordering operator in imaginary time $\tau$, and $\langle \ \cdot\ \rangle$ stands for the thermal average.
The Fourier representation for imaginary time is introduced as 
\begin{align}
  \mathcal{G}_{\bm{k},\eta\eta'}(i\omega_n) = \int_{0}^\beta d\tau e^{i\omega_n\tau} \mathcal{G}_{\bm{k},\eta\eta'}(\tau),
\end{align}
where $\omega_n=2n\pi/\beta$ is the Matsubara frequency with $n$ being integer, $k_{\rm B}$ is the Boltzmann constant, and $T$ is the temperature.

The bare Green's function is given by
\begin{align}
  \mathcal{G}_{\bm{k}}^{(0)} (i\omega_n) = \left[i\omega_n\sigma_3 - \mathcal{E}_{\bm{k}}\right]^{-1},
\end{align}
which is a $2N\times 2N$ matrix.
Moreover, the temperature Green's function can be expanded as
\begin{align}
  \label{eq:Green perturbation}
  \mathcal{G}_{\bm{k}} (\tau) = &
  \mathcal{G}_{\bm{k}}^{(0)} (\tau) + \int_{0}^\beta d\tau_1
  \Big\langle
  T_{\tau} \mathcal{H}_{\text{int}} (\tau_1) \mathcal{B}_{\bm{k}}^{}(\tau) \mathcal{B}_{\bm{k}}^\dagger
  \Big\rangle_{0}\nonumber\\
  &- \frac{1}{2!}
  \int_{0}^\beta d\tau_1 \int_{0}^\beta d\tau_2
  \Big\langle
  T_{\tau} \mathcal{H}_{\text{int}}(\tau_1) 
  \mathcal{H}_{\text{int}}(\tau_2)
  \mathcal{B}_{\bm{k}}^{}(\tau) \mathcal{B}_{\bm{k}}^\dagger
  \Big\rangle_{0}\nonumber\\
  &\qquad\qquad + \cdots,
\end{align}
where $\langle\ \cdot\ \rangle_{0}$ represents the thermal average for the unperturbed Hamiltonian $\mathcal{H}_{0}$.
We also introduce the self-energy as
\begin{align}
    \label{appeq:temperature self-energy}
    \Sigma_{\bm{k}} (i\omega_n) &\equiv  \Big[\mathcal{G}_{\bm{k}}^{(0)}(i\omega_n)\Big]^{-1} - \Big[\mathcal{G}_{\bm{k}} (i\omega_n)\Big]^{-1}.
\end{align}
The Green's function is written as $\mathcal{G}_{\bm{k}} (i\omega_n) = \Big[ i\omega_n\sigma_3 - \mathcal{E}_{\bm{k}} - \Sigma_{\bm{k}}(i\omega_n) \Big]^{-1}$.
The retarded and advanced self-energies, $\Sigma_{\bm{k}}^{R}(\omega)$ and $\Sigma_{\bm{k}}^{A}(\omega)$, are calculated by performing the analytic continuation.
By using the self-energy, the retarded, and advanced Green's functions can be written as~\cite{mahan}
\begin{align}
    {G}_{\bm{k}}^{R} (\omega) &=  \Big[ (\omega+i0^{+})\sigma_3 - \mathcal{E}_{\bm{k}} - \Sigma_{\bm{k}}^{R}(\omega) \Big]^{-1},\label{eq:interaction GreenR}\\
    {G}_{\bm{k}}^{A} (\omega) &= \Big[ (\omega-i0^{+})\sigma_3 - \mathcal{E}_{\bm{k}} - \Sigma_{\bm{k}}^{A}(\omega) \Big]^{-1}.
    \label{eq:interaction GreenA}
\end{align}
Note that the temperature Green's function satisfies the conditions $\mathcal{G}_{\bm{k},\eta\eta'} (i\omega_{n})= \mathcal{G}_{-\bm{k},\eta'+N,\eta+N} (-i\omega_{n})$ and $\mathcal{G}_{\bm{k},\eta,\eta'+N} (i\omega_{n}) = \mathcal{G}_{-\bm{k},\eta',\eta+N} (-i\omega_{n})$ for $0\leq \eta,\eta' \leq N$, which are obtained from Eq.~\eqref{eq:definition-Green}.
From them, the following relations hold for the retarded, and advanced Green's functions with $0\leq \eta,\eta' \leq N$:
\begin{align}
    G_{\bm{k},\eta\eta'}^{R} (\omega) &= G_{-\bm{k},\eta'+N,\eta+N}^{A} (-\omega),\\
    G_{\bm{k},\eta,\eta'+N}^{R} (\omega) &= G_{-\bm{k},\eta',\eta+N}^{A} (-\omega).
\end{align}
Similarly, the temperature, retarded, and advanced self energies satisfy the following relations for $0\leq \eta,\eta' \leq N$:
\begin{align}
    \Sigma_{\bm{k},\eta\eta'} (i\omega_{n}) &= \Sigma_{-\bm{k},\eta'+N,\eta+N} (-i\omega_{n})\\
    \Sigma_{\bm{k},\eta,\eta'+N} (i\omega_{n}) &= \Sigma_{-\bm{k},\eta',\eta+N} (-i\omega_{n}),
\end{align}
and
\begin{align}\label{eq:gamma-hp}
    \Sigma_{\bm{k},\eta\eta'}^{R} (\omega) &= \Sigma_{-\bm{k},\eta'+N,\eta+N}^{A} (-\omega)\\
    \Sigma_{\bm{k},\eta,\eta'+N}^{R} (\omega) &= \Sigma_{-\bm{k},\eta',\eta+N}^{A} (-\omega).
\end{align}

Finally, we introduce the spectral function as follows:
\begin{align}
    \rho_{\bm{k},\eta} (\omega) = \frac{1}{2\pi}\int_{-\infty}^{\infty} dt e^{i\omega t}\left\langle \left[ \mathcal{B}_{\bm{k},\eta}^{}(t), \mathcal{B}_{\bm{k},\eta}^\dagger \right] \right\rangle.
    \label{eq:spec-omega}
\end{align}
From Eq.~\eqref{eq:commu-B}, the sum rule is given by
\begin{align}
    \label{eq:sumrule-rho}
    \int_{-\infty}^{\infty} \rho_{\bm{k}} (\omega) d\omega = \sigma_{3}.
\end{align}
Furthermore, the spectral function is connected to the diagonal component of the retarded Green's function by the following relation: 
\begin{align}
    \label{eq:spec-Green}
    \rho_{\bm{k},\eta} (\omega) = -\frac{1}{\pi} \mathrm{Im} G^{R}_{\bm{k},\eta\eta} (\omega).
\end{align}
Meanwhile, the spectral function satisfies the condition $\mathrm{sgn} (\omega) \rho_{\bm{k},\eta}(\omega) \geq 0$, which is obtained from its Lehmann representation.
Thus, the following relation holds for the imaginary part of the retarded Green's function:
\begin{align}
    \mathrm{sgn}(\omega) \mathrm{Im} G^{R}_{\bm{k},\eta\eta} (\omega) &\leq 0.
    \label{eq:sign-ImGreen}
\end{align}

\section{Formalism for thermal transport}
\label{sec:thermal-transport}

In this section, we formulate thermal conductivity using the bosonic Green's function introduced in the previous section.

\subsection{Introduction to thermal conductivity}
\label{sec:intro-thermal-conductivity}

First, we briefly review a general theory for thermal transport based on Ref.~\cite{smrcka1977,cooper1997,xiao2006,qin2011}.
Thermal responses can be microscopically evaluated as a response against a virtually introduced gravitational field, which is the mechanical counterpart of the temperature gradient applied to the system~\cite{luttinger1964}.
Here, we consider the local Hamiltonian $h (\bm{r}_{i})$ involving site $i$, which satisfies $\mathcal{H} = \sum_{i}h (\bm{r}_{i})$.
From this local Hamiltonian, the external field is introduced by replacing $\mathcal{H}$ with $\mathcal{H}^{\chi} = \sum_{i}\left[1+\chi (\bm{r}_{i})\right] h(\bm{r}_{i})$, where $\chi (\bm{r})$ is the gravitational field applied to the system.
Since we consider a bosonic system with zero chemical potential, the thermal current operator is equivalent to the energy current, which is given by~\cite{mahan}
\begin{align}
    \label{eq:def-J}
    \bm{J}^{Q} = \frac{1}{V}\pdiff{\bm{P}_{E}}{t} = \frac{i}{V\hbar} \left[\mathcal{H},\bm{P}_E\right],
\end{align}
where $\bm{P}_E$ is the energy polarization operator defined as
\begin{align}
    \bm{P}_E = \sum_{i} \bm{r}_{i} h(\bm{r}_{i}).
\end{align}
In the presence of the gravitational field, the local Hamiltonian is replaced to $h(\bm{r}_{i})\to h^{\chi} (\bm{r}_{i}) = \left[1 + \chi (\bm{r}_{i})\right] h(\bm{r}_{i})$~\cite{luttinger1964,matsumoto2014}, and thermal current is also changed to  $\bm{J}^{Q;\chi}$, which is written as
\begin{align}
    \label{eq:def-J-chi}
    \bm{J}^{Q;\chi} = \frac{i}{V\hbar} \left[
    \mathcal{H}^{\chi}, 
    \bm{P}_E^{\chi}\right],
\end{align}
where $\bm{P}_{E}^{\chi} = \sum_{i} \bm{r}_{i} \left[1+\chi(\bm{r}_{i})\right] h(\bm{r}_{i}) = \sum_{i} \bm{r}_{i} h^{\chi} (\bm{r}_{i})$.

Here, we introduce the thermal conductivity $\kappa_{\lambda\lambda'}$, where $\lambda (=x,y,z)$ is the component of Cartesian coordinate, as
\begin{align}
    \label{eq:kxy-def}
    J^{\mathrm{tr}}_{\lambda} = \kappa_{\lambda\lambda'} (-\nabla_{\lambda'} T).
\end{align}
Here, $\nabla T$ is the temperature gradient, and $\bm{J}^{\mathrm{tr}}$ is the thermal transport response~\cite{smrcka1977,cooper1997,xiao2006,qin2011}.
Note that $\bm{J}^{\mathrm{tr}}$ must vanish in the absence of the thermal gradient.
As mentioned at the beginning of this section, the thermal conductivity can be evaluated as the response against the gradient of the gravitational field instead of the temperature gradient as follows~\cite{luttinger1964}:
\begin{align}
\label{eq:thermal coefficient-chi}
    J^{\mathrm{tr}}_{\lambda} = L_{\lambda\lambda'} \left(-\nabla_{\lambda'} \chi(\bm{r})\right),
\end{align}
where $L_{\lambda\lambda'}=\kappa_{\lambda\lambda'} T$ is the thermal transport coefficient with $T$ being the temperature in equilibrium.
The thermal transport response $\bm{J}^{\mathrm{tr}}$ is not equivalent to $\langle\bm{J}^{Q;\chi}\rangle_{\nabla \chi}$ in the first order of $\nabla \chi$, where $\braket{\ \cdot \ }_{\nabla \chi}$ is the expectation value in the presence of the gravitational-field gradient.
To enforce the condition of $\bm{J}^{\mathrm{tr}}=0$ for $\nabla \chi=0$, one needs to subtract the contribution from a heat magnetization from $\langle\bm{J}^{Q;\chi}\rangle_{\nabla \chi}$ (see Appendix~\ref{app:def-orbmag})~\cite{smrcka1977,cooper1997,xiao2006,qin2011}.
Finally, the thermal transport coefficient is written as
\begin{align}
    \label{eq:Lxy}
    L_{\lambda\lambda'} = S_{\lambda\lambda'} + 
    \sum_{\lambda''}\varepsilon_{\lambda\lambda'\lambda''}\frac{2M_{\lambda''}^{Q}}{V},
\end{align}
where $S_{\lambda\lambda'}$ is the contribution obtained by the well-known Kubo formula~\cite{kubo1957}, $\bm{M}^{Q}$ is the heat magnetization originating from thermal carriers~\cite{matsumoto2011,qin2011}, $V$ is the volume of the system, and $\varepsilon_{\lambda\lambda'\lambda''}$ is the Levi-Civita symbol.
Note that the second term in Eq.~\eqref{eq:Lxy} does not contribute to the symmetric components of the thermal transport coefficient but plays crucial role in the antisymmetric components, namely the thermal Hall conductivity.

The first term of Eq.~\eqref{eq:Lxy} is evaluated from 
\begin{align}
 \label{eq:def-Sxy}
    S_{\lambda\lambda'} = -\lim_{\Omega\to 0} \frac{P_{\lambda\lambda'}^{R} (\Omega) - P_{\lambda\lambda'}^{R}(\Omega)}{i\Omega},
\end{align}
where $P_{\lambda\lambda'}^{R}(\Omega)$ is the retarded correlation function between thermal currents, which is calculated from the imaginary-time correlation function $P_{\lambda\lambda'} (i\Omega)$ via analytic continuation: $P_{\lambda\lambda'}^{R}(\Omega) = P_{\lambda\lambda'} (i\Omega \to \hbar\Omega + i0^{+})$.
Here, $P_{\lambda\lambda'}(i\Omega)$ is given by
\begin{align}
    P_{\lambda\lambda'}(i\Omega) = - \frac{1}{V} \int_{0}^{\beta} d\tau e^{i\Omega\tau} 
    \langle T_{\tau} J_{\lambda}^{Q}(\tau) J_{\lambda'}^{Q} \rangle,
    \label{eq:Piw}
\end{align}
where $\beta = 1/k_{\rm B} T$ and the Heisenberg representation of an operator ${\cal O}$ is defined as
$
{\cal O} (\tau) = e^{\tau\mathcal{H}} {\cal O} e^{-\tau\mathcal{H}}
$.
On the other hand, the heat magnetization $\bm{M}^{Q}$ is evaluated from the following relations:
\begin{align}
    \label{eq:def-Mz}
    2 \bm{M}^{Q} + \beta \pdiff{\bm{M}^{Q}}{\beta} = \frac{1}{\beta} \pdiff{}{\beta}\left(\beta^2 \bm{M}^{Q}\right) = \tilde{\bm{M}}^{Q},
\end{align}
where $\tilde{\bm{M}}^{Q}$ is given by
\begin{align}
    \label{eq:def-tildeMz}
    \tilde{M}_{\lambda}^{Q} = -\frac{\beta}{2i} \sum_{\lambda'\lambda''}\varepsilon_{\lambda\lambda'\lambda''}\pdiff{}{q_{\lambda''}} \left\langle h_{-\bm{q}}^{};j_{\bm{q},\lambda'}^{Q}\right\rangle \Big|_{\bm{q}\to \bm{0}}.
\end{align}
The differential equation in Eq.~\eqref{eq:def-Mz} is solved under the boundary condition: $\lim_{\beta\to\infty}\beta \pdiff{\bm{M}^{Q}}{\beta} = 0$, namely, $2\bm{M}^{Q}=\tilde{\bm{M}}^{Q}$ at zero temperature limit~\cite{qin2011}.
Here, we introduce the Fourier transforms of the local Hamiltonian $h(\bm{r}_{i})$ and the thermal current density defined by $\bm{J}^{Q} = \sum_{i} \bm{j}^{Q}(\bm{r}_{i})$ as
\begin{align}
    h_{\bm{q}}= \sum_{i} h(\bm{r}_{i})e^{-i\bm{q}\cdot \bm{r}_{i}},\quad
    \bm{j}^{Q}_{\bm{q}}= \sum_{i} \bm{j}^{Q}(\bm{r}_{i}) e^{-i\bm{q}\cdot\bm{r}_{i}}.
\end{align}
Additionally, $\left\langle h_{-\bm{q}}^{};\bm{j}_{\bm{q}}^{Q}\right\rangle$ stands for the canonical correlation between them, which is defined by
\begin{align}
    \left\langle h_{-\bm{q}}^{};\bm{j}_{\bm{q}}^{Q}\right\rangle = \frac{1}{\beta} \int_{0}^{\beta} d\tau \left\langle h_{-\bm{q}}^{}(\tau) \bm{j}_{\bm{q}}^{Q}\right\rangle.
\end{align}

To evaluate heat magnetization from Eq.~\eqref{eq:def-Mz}, the following scaling relation must be imposed for the thermal current density~\cite{qin2011,cooper1997}:
\begin{align}
    \label{eq:scaling-j}
    \bm{j}^{Q;\chi}(\bm{r}_{i}) &= \left[1 + \chi(\bm{r}_{i}) \right]^2 \bm{j}^{Q} (\bm{r}_{i}).
\end{align}
In the following calculations, the local Hamiltonian and current density are introduced from the bilinear bosonic Hamiltonian given in Eq.~\eqref{eq:Hamil-A-real}, for simplicity.
Then, the local Hamiltonian is represented as
\begin{align}
    h_{\bm{q}} = \frac{1}{2} \sum_{\bm{k}}  \mathcal{A}_{\bm{k}}^\dagger \frac{\mathcal{M}_{\bm{k}}+\mathcal{M}_{\bm{k}+\bm{q}}}{2} \mathcal{A}_{\bm{k}+\bm{q}},
    \label{eq:hq-def}
\end{align}
and the thermal current density satisfying Eq.~\eqref{eq:scaling-j} is written as
\begin{align}
    \bm{j}^{Q}_{\bm{q}} =& 
    \frac{1}{4} \sum_{\bm{k}} 
    \mathcal{A}_{\bm{k}}^\dagger 
    \left( \bm{v}_{\bm{k}}^{} \sigma_3^{} \mathcal{M}_{\bm{k}+\bm{q}}^{} + \mathcal{M}_{\bm{k}}^{} \sigma_3^{} \bm{v}_{\bm{k}+\bm{q}}^{} \right) \mathcal{A}_{\bm{k}+\bm{q}}^{}\nonumber\\
    &-\frac{1}{16}\sum_{\lambda}\sum_{\bm{k}} \hbar q_{\lambda} 
    \mathcal{A}_{\bm{k}}^\dagger
    \left( \bm{v}_{\bm{k}}^{}\sigma_3^{} v_{\bm{k}+\bm{q},\lambda} - v_{\bm{k},\lambda}\sigma_3^{}\bm{v}_{\bm{k}+\bm{q}}^{} \right)
    \mathcal{A}_{\bm{k}+\bm{q}}^{},    \label{eq:scaling-jq}
\end{align}
where $\bm{v}_{\bm{q}} =\frac{1}{\hbar}\pdiff{\mathcal{M}_{\bm{q}}}{\bm{q}}$.
The derivation of the above representation is given in Appendix~\ref{app:tco and sl}.
Note that the total thermal current $\bm{J}^{Q}$ is given by $\bm{J}^{Q}=\bm{j}^{Q}_{\bm{q}}\bigr|_{\bm{q}\to 0}$, where the second term in Eq.~\eqref{eq:scaling-jq} does not contribute to $\bm{J}^{Q}$, and thereby, $\bm{J}^{Q}$ is written as
\begin{align}
    \bm{J}^{Q}= 
    \frac{1}{4} \sum_{\bm{k}} 
    \mathcal{A}_{\bm{k}}^\dagger 
    \left( \bm{v}_{\bm{k}}^{} \sigma_3^{} \mathcal{M}_{\bm{k}}^{} + \mathcal{M}_{\bm{k}}^{} \sigma_3^{} \bm{v}_{\bm{k}}^{} \right) \mathcal{A}_{\bm{k}}^{}.
\label{eq:JQ-form}
\end{align}

\subsection{Green's function representation of thermal conductivity}
\label{sec:Sxy}

In this section, we show the representations of $S_{\lambda\lambda'}$ and $\tilde{\bm{M}}^{Q}$ using the Green's functions introduced in Eqs.~\eqref{eq:interaction GreenR} and \eqref{eq:interaction GreenA}, where we neglect vertex corrections (see Appendix~\ref{app:transport-coefficient}).
For simplicity, we omit their off-diagonal components with respect to $\eta$ and only consider diagonal components of the retarded and advanced Green's functions, which are written as $G_{\bm{k},\eta}^{R}(\omega)$ and $G_{\bm{k},\eta}^{A}(\omega)$, respectively.
Under this assumption, $S_{\lambda\lambda'}$ and $\tilde{M}_{\lambda}^{Q}$ are represented as
\begin{widetext}
\begin{align}
    S_{\lambda\lambda'} = &-\frac{i\hbar}{8V} \sum_{\eta,\eta'=1}^{2N} \sum_{\bm{k}}
    \left( T_{\bm{k}}^\dagger v_{\bm{k},\lambda}^{} T_{\bm{k}}^{} \right)_{\eta\eta'} \left( T_{\bm{k}}^\dagger v_{\bm{k},\lambda'}^{} T_{\bm{k}}^{} \right)_{\eta'\eta} 
    \left( \sigma_{3,\eta}\mathcal{E}_{\bm{k},\eta} + \sigma_{3,\eta'}\mathcal{E}_{\bm{k},\eta'} \right)^2\nonumber\\
    &\qquad\qquad\qquad\times {\rm P}\int_{-\infty}^{\infty} \frac{d\omega}{\pi} g(\beta\omega)
    \left\{
    \mathrm{Im} \left[G_{\bm{k},\eta}^{R}(\omega)\right] \pdiff{G_{\bm{k},\eta'}^{R}(\omega)}{\omega}
    -
    \pdiff{G_{\bm{k},\eta}^{A}(\omega)}{\omega} \mathrm{Im} \left[G_{\bm{k},\eta'}^{R} (\omega)\right] 
    \right\},
    \label{eq:Sxy-green}
\end{align}
and
\begin{align}
    \tilde{M}_{\lambda}^{Q}
    =  
    &-\frac{1}{16i}\sum_{\lambda'\lambda''}\varepsilon_{\lambda\lambda'\lambda''} \pdiff{}{q_{\lambda''}} \sum_{\eta,\eta'=1}^{2N} \sum_{\bm{k}} 
    \left[T_{\bm{k}}^\dagger \left(\mathcal{M}_{\bm{k}}+\mathcal{M}_{\bm{k}-\bm{q}}\right) T_{\bm{k}-\bm{q}}^{}\right]_{\eta\eta'}
        \nonumber\\
    &\qquad\times\Bigg[
    T_{\bm{k}-\bm{q}}^\dagger \left(
     v_{\bm{k}-\bm{q},\lambda}^{} \sigma_3^{} \mathcal{M}_{\bm{k}}^{} + \mathcal{M}_{\bm{k}-\bm{q}}^{} \sigma_3^{} v_{\bm{k},\lambda}^{}
    - \sum_{\lambda'''} \hbar q_{\lambda'''}^{}
    \frac{v_{\bm{k}-\bm{q},\lambda}^{} \sigma_3^{} v_{\bm{k},\lambda'''}^{} - v_{\bm{k}-\bm{q},\lambda'''}^{} \sigma_3^{} v_{\bm{k},\lambda}^{}}{4} \right)
    T_{\bm{k}}^{} 
    \Bigg]_{\eta'\eta}
    \nonumber\\
    & \qquad\qquad\qquad\qquad
    \times\mathrm{P}\int_{-\infty}^{\infty} \frac{d\omega}{\pi}  g(\beta\omega)
    \Bigg\{
      \mathrm{Im} \left[G_{\bm{k},\eta}^{R} (\omega)\right] G_{\bm{k-\bm{q}},\eta'}^{R}(\omega)
      + G_{\bm{k},\eta}^{A}(\omega) \mathrm{Im}\left[G_{\bm{k}-\bm{q},\eta'}^{R}(\omega)\right]
    \Bigg\}\Bigg|_{\bm{q}\to 0},
    \label{eq:tildeMz-green}
\end{align}
\end{widetext}
respectively.
Here, $g(x) = (e^{x} - 1)^{-1}$ is the Bose distribution function with zero chemical potential, and ${\rm P}\int$ stands for the principal value integral.
The details of the derivations for Eqs.~\eqref{eq:Sxy-green} and \eqref{eq:tildeMz-green} are given in Appendices~\ref{app:express-Sxy} and \ref{app:express-tildeMz}, respectively.

\subsection{Thermal Hall conductivity with approximate Green's function}
\label{sec:appro-green}

Here, we rewrite $S_{\lambda\lambda'}$ and $\tilde{M}_{\lambda}^{Q}$ given in Eqs.~\eqref{eq:Sxy-green} and \eqref{eq:tildeMz-green}, respectively, as more convenient expressions.
In the present study, we focus on effects of magnon damping on the thermal Hall effect, and hence, we take into account the imaginary part of the self-energy in Eq.~\eqref{eq:interaction GreenR} and \eqref{eq:interaction GreenA} and neglect its real part.
In the previous section, we only consider the diagonal part of the Green's functions.
This simplification corresponds to omitting the off-diagonal components of the self-energy, and hence, we here only consider the imaginary part of the diagonal components of the self-energy.
We define the imaginary part of the retarded self-energy as
\begin{align}
    \label{eq:gamma-def}
    \Gamma_{\bm{k},\eta} (\omega) = - \mathrm{Im} \Sigma_{\bm{k},\eta}^{R} (\omega),
\end{align}
which corresponds to the damping rate of the magnon with momentum $\hbar\bm{k}$ and branch $\eta$.
Using the above approximation and the damping rate $\Gamma_{\bm{k},\eta} (\omega)$, we represent the retarded and advanced Green's functions as
\begin{align}
    \label{eq:interaction GreenR-appro}
    {G}_{\bm{k},\eta}^{R} (\omega) &\simeq  \frac{1}{(\omega+i0^{+})\sigma_{3,\eta} - \mathcal{E}_{\bm{k},\eta} + i\Gamma_{\bm{k},\eta}(\omega)},\\
    \label{eq:interaction GreenA-appro}
    {G}_{\bm{k},\eta}^{A} (\omega) &\simeq  \frac{1}{(\omega-i0^{+})\sigma_{3,\eta} - \mathcal{E}_{\bm{k},\eta} - i\Gamma_{\bm{k},\eta}(\omega)},
\end{align}
where we use the relation $\mathrm{Im}\Sigma_{\bm{k},\eta}^{R} (\omega)=-\mathrm{Im}\Sigma_{\bm{k},\eta}^{A} (\omega)$, which is obtained from the Lehmann representation of the Green's functions~\cite{mahan}.

Here, we discuss the analytical properties of $\Gamma_{\bm{k},\eta}(\omega)$.
From Eq.~\eqref{eq:sign-ImGreen}, the retarded self-energy satisfies $\mathrm{sgn}(\omega) \mathrm{Im}\Sigma_{\bm{k},\eta}^{R}(\omega) \leq 0$.
This leads to the following conditions:
\begin{align}
    \mathrm{sgn}(\omega)\Gamma_{\bm{k},\eta} (\omega) \geq 0.\label{eq:Gamma-positive}
\end{align}
Additionally, from Eq.~\eqref{eq:gamma-hp}, $\Gamma_{\bm{k},\eta}(\omega)$ also satisfies the following relation for $\eta = 1,\cdots, N$:
\begin{align}
    \Gamma_{\bm{k},\eta}(\omega) = - \Gamma_{-\bm{k},\eta+N} (-\omega).
    \label{eq:Gamma-eta}
\end{align}

Hereafter, we assume that $\Gamma_{\bm{k}}(\omega)$ varies slowly enough as a function of $\bm{k}$, $\omega$, and $T$.
We neglect the differential coefficients with respect to these variables and focus only on the antisymmetric part of the thermal conductivity matrix $\kappa_{\lambda\lambda'}$ to discuss the thermal Hall effect.
Within the assumption, the $\omega$ derivative of the retarded and advanced Green's functions are represented as
$\partial G_{\bm{k},\eta}^{R}/\partial\omega \simeq - 1/[\omega\sigma_{3,\eta} - \mathcal{E}_{\bm{k},\eta} + i\Gamma_{\bm{k},\eta}(\omega)]^2$ and $\partial G_{\bm{k},\eta}^{A}/\partial\omega \simeq -1/[\omega\sigma_{3,\eta} - \mathcal{E}_{\bm{k},\eta} - i\Gamma_{\bm{k},\eta}(\omega)]^2$, respectively.
Using the representations of Green's functions in Eqs.~\eqref{eq:interaction GreenR-appro} and \eqref{eq:interaction GreenA-appro} and the above approximations, the antisymmetric part of $S_{\lambda\lambda'}$ and $2M_{\lambda}^{Q}/V$ are calculated as
\begin{widetext}
\begin{align}
    \label{eq:Sxy}
    S_{\lambda\lambda'}^{a} &\simeq
\frac{1}{4\hbar V} \sum_{\lambda''}\varepsilon_{\lambda\lambda'\lambda''}
    \sum_{\eta=1}^{N}\sum_{\eta'=1}^{2N}\sum_{\bm{k}} \tilde{\Omega}_{\bm{k},\eta\eta'}^{\lambda''}
    \left( \varepsilon_{\bm{k},\eta} + \sigma_{3,\eta'}{\mathcal{E}}_{\bm{k},\eta'} \right)^2 \left( \varepsilon_{\bm{k},\eta} - \sigma_{3,\eta'}\mathcal{E}_{\bm{k},\eta'} \right)^2
            \nonumber\\
    &\qquad\qquad\qquad\times
    \mathrm{Re}
    \left[
       \mathrm{P}\int_{-\infty}^{\infty} d\omega
        \rho_{\bm{k},\eta} (\omega) \frac{2g(\beta\omega) + 1}{\left(\omega - \sigma_{3,\eta'}\mathcal{E}_{\bm{k},\eta'}+i\sigma_{3,\eta'}\Gamma_{\bm{k},\eta'}(\omega)\right)^2}
    \right]
\end{align}
and
\begin{align}
    \frac{2M_{\lambda}^{Q}}{V}
    \simeq &
    -\frac{1}{2\beta^{2}\hbar V} \sum_{\eta=1}^{N}\sum_{\eta'=1}^{2N}\sum_{\bm{k}} \tilde{\Omega}_{\bm{k},\eta\eta'}^{\lambda}
    \left( \varepsilon_{\bm{k},\eta} + \sigma_{3,\eta'}\mathcal{E}_{\bm{k},\eta'} \right)^2
    \left( \varepsilon_{\bm{k},\eta} - \sigma_{3,\eta'}\mathcal{E}_{\bm{k},\eta'}\right)
        \nonumber\\
    &\qquad\qquad\qquad\times
    \mathrm{Re}
    \left[\mathrm{P}
    \int_{-\infty}^{\infty} d\omega
        \frac{\rho_{\bm{k},\eta}(\omega)}{\omega-\sigma_{3,\eta'}\mathcal{E}_{\bm{k},\eta'} + i\sigma_{3,\eta'}\Gamma_{\bm{k},\eta'}(\omega)}\right]
    \int_{0}^{\beta}
    \tilde{\beta} \left[2g(\tilde{\beta}\omega) + 1 \right] d\tilde{\beta}
    \nonumber\\
    &-\frac{1}{4\beta^{2}\hbar V} \sum_{\eta=1}^{N}\sum_{\eta'=1}^{2N} \sum_{\bm{k}} 2\tilde{\Omega}_{\bm{k},\eta\eta'}^\lambda 
    \varepsilon_{\bm{k},\eta}
    \left[ \left( \varepsilon_{\bm{k},\eta} +\sigma_{3,\eta'}\mathcal{E}_{\bm{k},\eta'} \right)^2 - 4 \varepsilon_{\bm{k},\eta}^2 \right]
    \mathrm{P}
    \int_{-\infty}^{\infty}
    d\omega \rho_{\bm{k},\eta}(\omega)
    \int_{0}^{\beta}
    \tilde{\beta} \pdiff{g}{\omega} d\tilde{\beta},
    \label{eq:Mz}
\end{align}
\end{widetext}
where $S_{\lambda\lambda'}^{a}=(S_{\lambda\lambda'}-S_{\lambda'\lambda})/2$
and 
$\tilde{\Omega}_{\bm{k},\eta\eta'}^\lambda$ are defined as
\begin{align}
    \tilde{\Omega}_{\bm{k},\eta\eta'}^\lambda &=
    -\frac{i\hbar^2}{2} 
    \sum_{\lambda'\lambda''}
    \varepsilon_{\lambda\lambda'\lambda''}
    \frac{
    \sigma_{3,\eta}\sigma_{3,\eta'} 
    \left( T_{\bm{k}}^\dagger v_{\bm{k},\lambda'}^{} T_{\bm{k}}^{} \right)_{\eta\eta'} \left( T_{\bm{k}}^\dagger v_{\bm{k},\lambda''}^{} T_{\bm{k}}^{} \right)_{\eta'\eta}
    }
    {\left( \sigma_{3,\eta}\mathcal{E}_{\bm{k},\eta} - \sigma_{3,\eta'}\mathcal{E}_{\bm{k},\eta'} \right)^2} 
    \label{eq:Omega-tilde}
\end{align}
The detailed derivations of Eqs.~\eqref{eq:Sxy} and \eqref{eq:Mz} are given in Appendixes~\ref{app:calc-Sxy} and~\ref{app:calc-Mz}.
From Eq.~\eqref{eq:Lxy}, we find that the thermal Hall conductivity $\kappa_{\lambda\lambda'}^H=(\kappa_{\lambda\lambda'}-\kappa_{\lambda'\lambda})/2$ is evaluated by
\begin{align}
    \kappa_{\lambda\lambda'}^H=\frac{L_{\lambda\lambda'}^a}{T_0} = S_{\lambda\lambda'}^a + 
    \sum_{\lambda''}\varepsilon_{\lambda\lambda'\lambda''}\frac{2M_{\lambda''}^{Q}}{V},
\end{align}
where $L_{\lambda\lambda'}^a=(L_{\lambda\lambda'}-L_{\lambda'\lambda})/2$.

To further proceed the calculations of the thermal Hall conductivity, 
we presume that $\rho_{\bm{k},\eta} (\omega)$ for $\omega\to 0$ is sufficiently smaller than the maximum of $\rho_{\bm{k},\eta} (\omega)$.
This assumption is justified when the damping rate of magnons is small enough in the vicinity of the zero energy.
Previous studies have suggested that this situation is realized at low temperatures~\cite{chernyshev2009,mourigal2010,mook2020,koyama2023}.
Then, the contributions around the pole of the Green's function, $\omega\simeq \varepsilon_{\bm{k},\eta}$, are dominant in the evaluations of Eq.~\eqref{eq:Sxy} and \eqref{eq:Mz} for $\eta = 1,2,\cdots, N$.
Moreover, we hypothesize that $\Gamma_{\bm{k},\eta}(\omega)$ is small and incorporate this contribution up to the first order.
Based on the above assumptions, we obtain the representation of the thermal Hall conductivity as
\begin{align}
    \label{eq:kxy-A}
    \kappa_{\lambda\lambda'}^H
    \simeq & - \frac{k_{\rm B}^2 T}{\hbar V}\sum_{\lambda''}
    \sum_{\eta=1}^{N}\sum_{\bm{k}} \varepsilon_{\lambda\lambda'\lambda''}\Omega_{\bm{k},\eta}^{\lambda''}
    \notag\\
    &\qquad\times
    \int_{-\infty}^{\infty} d\omega
    \rho_{\bm{k},\eta}(\omega) c_2(g(\beta\omega)).
\end{align}
The derivation of the above expression is given in Appendix~\ref{app:calc-kxy}.
Here, we introduce $c_{2}(x)$ as
\begin{align}
    c_{2}(x) = \int_{0}^{x} \left( \ln \frac{1+t}{t} \right)^2 dt
\end{align}
and $\Omega_{\bm{k},\eta}$ is the Berry curvature given by~\cite{shindou2013,matsumoto2014}
\begin{align}
    \label{eq:def-berry}
    \Omega_{\bm{k},\eta}^\lambda = i \sum_{\lambda'\lambda''}\varepsilon_{\lambda\lambda'\lambda''}\left[ \sigma_3 \pdiff{T_{\bm{k}}^\dagger}{k_{\lambda'}} \sigma_3 \pdiff{T_{\bm{k}}^{}}{k_{\lambda''}} \right]_{\eta\eta} .
\end{align}
This Berry curvature satisfies the sum rule $\sum_{\eta=1}^{N} \sum_{\bm{k}}\Omega_{\bm{k},\eta}^\lambda = 0$ for the positive-energy branches~\cite{shindou2013,matsumoto2014} and is related to $\tilde{\Omega}_{\bm{k},\eta\eta'}^\lambda$ in Eq.~\eqref{eq:Omega-tilde} as $\Omega_{\bm{k},\eta}^\lambda= -2 \sum_{\eta'(\neq\eta)}^{2N}
\tilde{\Omega}_{\bm{k},\eta\eta'}^\lambda$ [see Eq.~\eqref{appeq:berry-trans}].

Here, we briefly discuss the noninteracting limit with $\Gamma_{\bm{k},\eta} (\omega) = 0$.
In this limit, $\rho_{\bm{k},\eta}(\omega) \to \delta(\omega-\varepsilon_{\bm{k},\eta})$ for $\omega\geq 0$, and hence, $\kappa_{xy}^{H}$ is written as
\begin{align}
    \label{eq:kxy-free}
    \kappa_{\lambda\lambda'}^{H;\mathrm{free}} 
    = -\frac{k_{\rm B}^2 T}{\hbar V} \sum_{\lambda''}\sum_{\eta=1}^{N} \sum_{\bm{k}} \varepsilon_{\lambda\lambda'\lambda''}\Omega_{\bm{k},\eta}^{\lambda''}
    c_2 (g(\beta\varepsilon_{\bm{k},\eta})).
\end{align}
This result coincides with the previous studies on free magnon systems~\cite{matsumoto2011,shindou2013,matsumoto2014}.

In numerical calculations, it is considerably difficult to evaluate the $\omega$ dependence of $\Gamma_{\bm{k},\eta}(\omega)$.
To reduce the calculation cost, we omit the $\omega$ dependence from $\Gamma_{\bm{k},\eta}(\omega)$ as $\tilde{\Gamma}_{\bm{k},\eta}$ for $\eta = 1,2, \cdots, N$.
The damping rate $\tilde{\Gamma}_{\bm{k},\eta}$ can be obtained by numerical calculations such as on-shell approximation as $\tilde{\Gamma}_{\bm{k},\eta}= \Gamma_{\bm{k},\eta}(\varepsilon_{\bm{k},\eta})$~\cite{harris1971,zhitomirsky2013} or off-shell methods~\cite{zhitomirsky2013,chernyshev2009,maksimov2016_prb,winter2017_nc,koyama2023}.
From Eq.~\eqref{eq:Gamma-positive}, we find that $\tilde{\Gamma}_{\bm{k},\eta}\geq 0$ because $\varepsilon_{\bm{k},\eta}$ is positive.
As an approximate form of $\Gamma_{\bm{k},\eta}(\omega)$ to satisfy these conditions and Eq.~\eqref{eq:Gamma-positive}, we introduce the following expression:
\begin{align}
    \Gamma_{\bm{k},\eta}(\omega)\simeq \tilde{\Gamma}_{\bm{k},\eta}\theta(\omega) \quad (\eta \leq N),
    \label{eq:def-Gamma-tilde}
\end{align}
where $\theta(\omega)$ is the step function.
In this approximation, the retarded Green's function is simplified as
\begin{align}
    G^{R}_{\bm{k},\eta} (\omega) \simeq \frac{1}{\omega+i0^{+} -  \varepsilon_{\bm{k},\eta} + i \tilde{\Gamma}_{\bm{k},\eta} \theta(\omega)}
    \quad (\eta \leq N).
\end{align}
This expression obviously satisfies Eq.~\eqref{eq:sign-ImGreen}.
Furthermore, within this approximation, $\rho_{\bm{k},\eta}(\omega)$ is written as $\rho_{\bm{k},\eta}(\omega) \simeq \tilde{\rho}_{\bm{k},\eta}(\omega)=\mathcal{L}_{\bm{k},\eta}(\omega)\theta(\omega)$, where $\mathcal{L}_{\bm{k},\eta}(\omega)$ is the Lorentz function given by
\begin{align}
    \mathcal{L}_{\bm{k},\eta}(\omega) = \frac{\tilde{\Gamma}_{\bm{k},\eta}/\pi}{(\omega - \varepsilon_{\bm{k},\eta})^2 + \tilde{\Gamma}_{\bm{k},\eta}^2} \quad (\eta \leq N).
\end{align}
Note that $I_{\bm{k},\eta}=\int_{-\infty}^\infty \tilde{\rho}_{\bm{k},\eta} (\omega)$ does not coincide with unity due to the simplification given in Eq.~\eqref{eq:def-Gamma-tilde}.
This quantity is written as $I_{\bm{k},\eta}= \frac12 + \frac1\pi\arctan \frac{\varepsilon_{\bm{k},\eta}}{\tilde{\Gamma}_{\bm{k},\eta}}$,
indicating that $I_{\bm{k},\eta}\simeq 1$ for $\tilde{\Gamma}_{\bm{k},\eta}\ll \varepsilon_{\bm{k},\eta}$.
Thus sum rule for $\tilde{\rho}_{\bm{k},\eta} (\omega)$ is approximately satisfied as long as $\tilde{\rho}_{\bm{k},\eta} (\omega)$ for $\omega\to 0$ is sufficiently small, which is assumed before.
Finally, we obtain the thermal Hall conductivity incorporating the magnon damping $\tilde{\Gamma}_{\bm{k},\eta}$ as
\begin{align}
    \label{eq:kxy-L}
    \kappa_{\lambda\lambda'}^{H} \simeq &-\frac{k_{B}^2 T}{\hbar V}\sum_{\lambda''}\sum_{\eta}^{N} \sum_{\bm{k}} \varepsilon_{\lambda\lambda'\lambda''}\Omega_{\bm{k},\eta}^{\lambda''}\nonumber\\
    &\qquad\times
    \int_{-\infty}^{\infty}
    d\omega \tilde{\rho}_{\bm{k},\eta}(\omega)  c_2(g(\beta\omega)).
\end{align}
In the following, we focus on $\kappa_{xy}^{H}$ in two-dimensional systems stacked along the $z$ direction, where the interlayer distance is assumed to be unity.

\subsection{Fundamental properties of thermal Hall conductivity incorporating magnon damping}
\label{sec:thermal-Hall-damping}

\begin{figure}[t]
  \begin{center}
    \includegraphics[width=\columnwidth,clip]{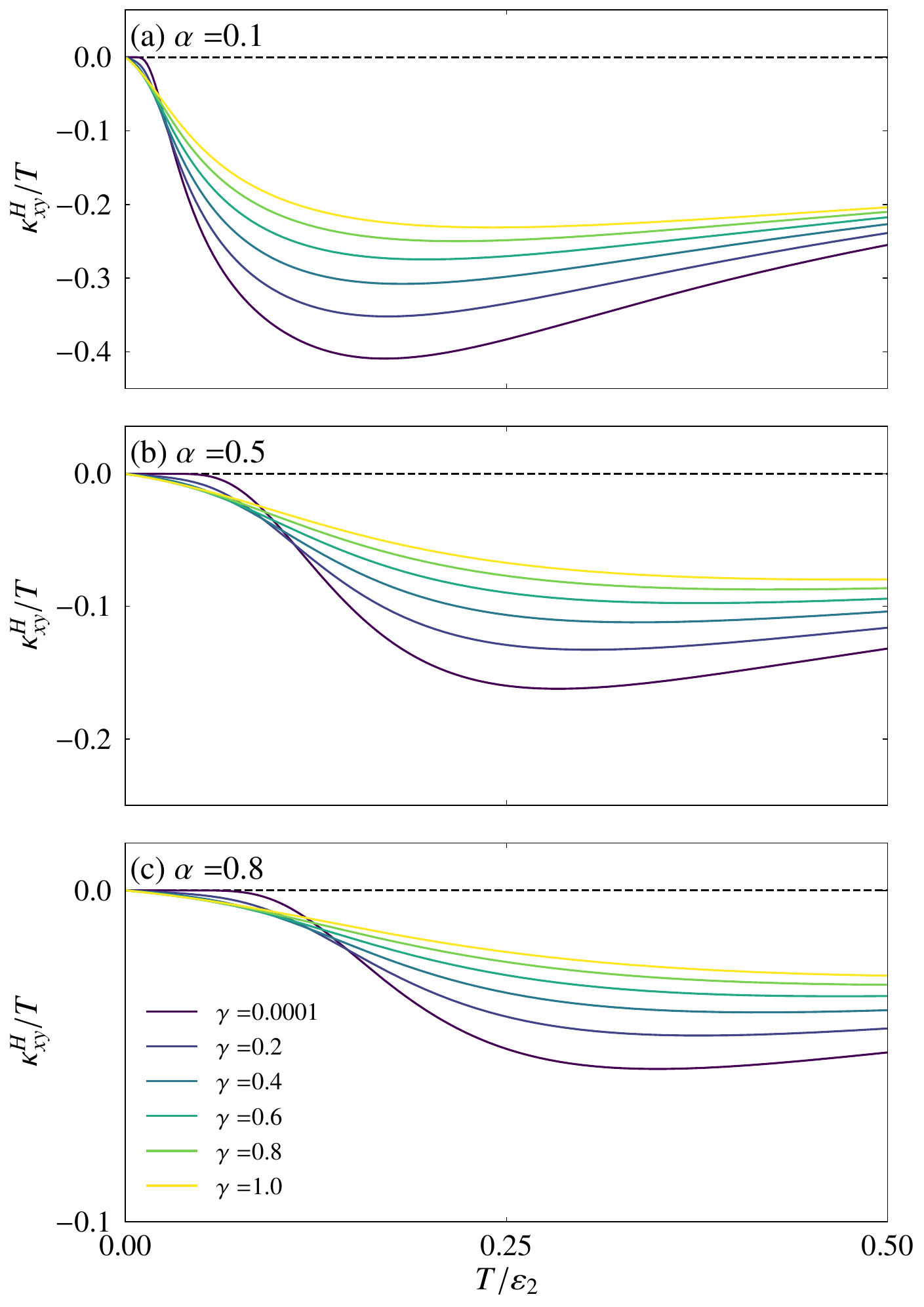}
    \caption{
    Temperature dependence of the thermal Hall conductivity in the two-band magnon model introduced in Sec.~\ref{sec:thermal-Hall-damping} for several values of the energy ratio $\alpha=\varepsilon_{\bm{k},1}/\varepsilon_{\bm{k},2}$.
    }
    \label{fig:kxy}
  \end{center}
\end{figure}

In the previous section, we have formulated the thermal Hall conductivity incorporating magnon damping as Eq.~\eqref{eq:kxy-L}.
In this section, we examine the fundamental properties by introducing a simple two-band magnon model ($N=2$), where the magnon dispersions and the corresponding damping rates are given by $\varepsilon_{\bm{k},\eta}$ and $\tilde{\Gamma}_{\bm{k},\eta}$ with $\eta=1,2$, respectively. 
Here, we omit the $\bm{k}$ dependence of the magnon energy and the damping rate and introduce the parameters $\alpha$ and $\gamma$ as $\alpha=\varepsilon_{\bm{k},1}/\varepsilon_{\bm{k},2}$ and $\gamma=\tilde{\Gamma}_{\bm{k},1}/\varepsilon_{\bm{k},1}=\tilde{\Gamma}_{\bm{k},2}/\varepsilon_{\bm{k},2}$, respectively.
We assume that $\varepsilon_{\bm{k},1}<\varepsilon_{\bm{k},2}$, namely, $\alpha<1$.
Furthermore, the Chern numbers $\mathcal{C}_{\eta}^{z}$ of the two magnon branches are set to be $\mathcal{C}_{1}^{z}=-\mathcal{C}_{2}^{z}=1$ where $\mathcal{C}_{\eta}^{z} = \frac{1}{2\pi}\int_{\mathrm{BZ}} dk_{x}dk_{y} \Omega_{\bm{k},\eta}^{\lambda}$.
The thermal Hall conductivity in this simple model is written as
\begin{align}
    \label{eq:kxy-nok}
    \kappa_{xy}^{H} 
    \simeq
    -\frac{k_{\rm B}^2 T}{2\pi\hbar} 
    \int_{-\infty}^{\infty}
    d\omega\left[ \tilde{\rho}_{1}(\omega)-\tilde{\rho}_{2}(\omega) \right]c_2(g(\beta\omega)).
\end{align}
Note that the temperature dependence of $\kappa_{xy}^{H}/T$ comes from $c_2(g(\beta\omega))$, and $c_2(g(x))$ is a monotonically decreasing function of $x$ from $\pi^2/3$ at $x=0$ to 0 at $x\to \infty$.

Figure~\ref{fig:kxy} shows the temperature dependence of $\kappa_{xy}^{H}/T$ in the simple two-band model for several values of $\alpha$.
As an overall behavior regardless of $\alpha$, we find that introducing the magnon damping enhances the absolute value of $\kappa_{xy}^{H}$ in the low-temperature region and suppresses it in the high-temperature region.
The impact of the magnon damping on the thermal Hall conductivity is understood as follows.
At low temperatures, the $\omega$ dependence of $c_2(g(\beta\omega))$ predominantly enhances the contribution of the low-energy part of the integral in Eq.~\eqref{eq:kxy-nok} to the thermal Hall conductivity.
This is amplified by magnon damping, which increases the lower-energy spectral weight, thereby enhancing $\kappa_{xy}^{H}/T$.
On the other hand, at higher temperatures, the $\omega$ dependence of $c_2(g(\beta\omega))$ becomes less pronounced.
Here, the magnon damping leads to a broadening of the spectrum.
This broadening facilitates an cancellation effect between contributions from the bands possessing the opposite Chern numbers, resulting in a reduction of the absolute value of $\kappa_{xy}^{H}/T$.
This effect is particularly pronounced when $\alpha$ is large, as shown in Fig.~\ref{fig:kxy}(c), because of the proximity of the two branches.
In contrast, at a smaller $\alpha$, the temperature range over which $\kappa_{xy}^{H}/T$ is enhanced by $\gamma$ becomes more restricted, as shown in Fig.~\ref{fig:kxy}(a).

\section{Application to localized systems}
\label{sec:result}

In this section, we apply our theory for the thermal Hall conductivity formulated above to the two localized spin-1/2 models on a honeycomb lattice: the Kitaev model under magnetic fields and the Heisenberg-DM model.
We evaluate the damping rate $\tilde{\Gamma}_{\bm{k},\eta}$ for each magnon branch $\eta$ using the self-consistent imaginary Dyson equation (iDE) approach at finite temperatures developed in Ref.~\cite{koyama2023}.
In this method, we consider contributions up to $O(1/\mathcal{S})$ corrections from the bilinear term ${\cal H}_0$ in Eq.~\eqref{eq:bosonic_H}.
To this end, we deal with $\mathcal{H}_3/\sqrt{\mathcal{S}}$ up to second-order perturbations and $\mathcal{H}_4/\mathcal{S}$ up to first-order perturbations~\cite{chernyshev2009,mook2020}.
We calculate $\kappa_{xy}^H$ based on Eq.\eqref{eq:kxy-L}, where the honeycomb lattice is defined on the $xy$ plane.

\subsection{Kitaev model under magnetic field}
\label{sec:Kitaev}

\begin{figure}[t]
  \begin{center}
    \includegraphics[width=\columnwidth,clip]{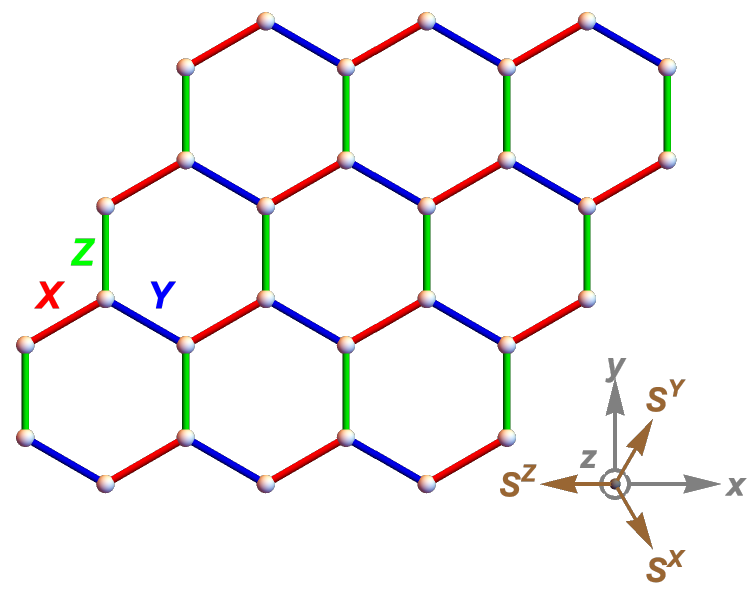}
    \caption{
    Schematic picture of the honeycomb lattice on which the Kitaev model is defined.
    The red, blue, and green lines represent the $X$, $Y$, and $Z$ bonds, respectively.
    The inset shows the relation between the coordinate of the spin space $(S^X,S^Y,S^Z)$ and that of the real space $(x,y,z)$.
    }
    \label{fig:honeycomb_kitaev}
  \end{center}
\end{figure}

First, we address thermal transport in the Kitaev model on a honeycomb lattice under an external magnetic field, whose Hamiltonian is given as follows:
\begin{align}
    \label{eq:kitaev-model}
    \mathcal{H} = 2K\sum_{\braket{ij}_{\varLambda}} S_{i}^\varLambda S_{j}^\varLambda - \sum_{i}\bm{h}\cdot\bm{S}_{i},
\end{align}
where $K$ and $\bm{h}$ are the strengths of the Kitaev interaction and magnetic field, respectively, and $S_{i}^\varLambda$ ($\varLambda=X,Y,Z$) is the $\varLambda$ component of an $S=1/2$ spin defined at site $i$ on the honeycomb lattice, whose bonds are classified into three types: $X$, $Y$, and $Z$ bonds, as shown in Fig.~\ref{fig:honeycomb_kitaev}.
The $\varLambda$ bond connecting between sites $i$ and $j$ are denoted as $\means{ij}_{\varLambda}$.
In this system, the $[111]$ axis of the spin space is taken to be parallel to the $z$ direction in the real space for the correspondence to real materials with spin-orbit coupling.
The other axes are determined as presented in the inset of Fig.~\ref{fig:honeycomb_kitaev}.
Here, we consider the ferromagnetic Kitaev model with $K<0$.
under a magentic field along the $[111]$ axis in the spin space, which corresponds to the out-of-plane $z$ direction.
We assume a forced ferromagnetic state along the external-field direction as a classical ground state.
Within the linear spin-wave approximation, gapped two magnon modes appear in the presence of the magnetic fields, and they exhibit nonzero nonzero Chern numbers with $\pm 1$~\cite{mcclarty2018,joshi2018}, leading to nonzero thermal Hall conductivity.
Note that the Hamiltonian does not commute with the total spin operator $\sum_{i} \tilde{S}_{i}^{Z}$, where $\tilde{S}_{i}^{Z}=(S_{i}^{X}+S_{i}^{Y}+S_{i}^{Z})/\sqrt{3}$ is the spin component along the field direction.
This suggests the appearance of magnon scattering processes without the particle number conservation, and thereby, the magnon damping should arise even in lower-order corrections for $1/{\cal S}$ in the Holstein-Primakoff theory~\cite{zhitomirsky2013,koyama2023}.

\begin{figure}[t]
  \begin{center}
    \includegraphics[width=\columnwidth,clip]{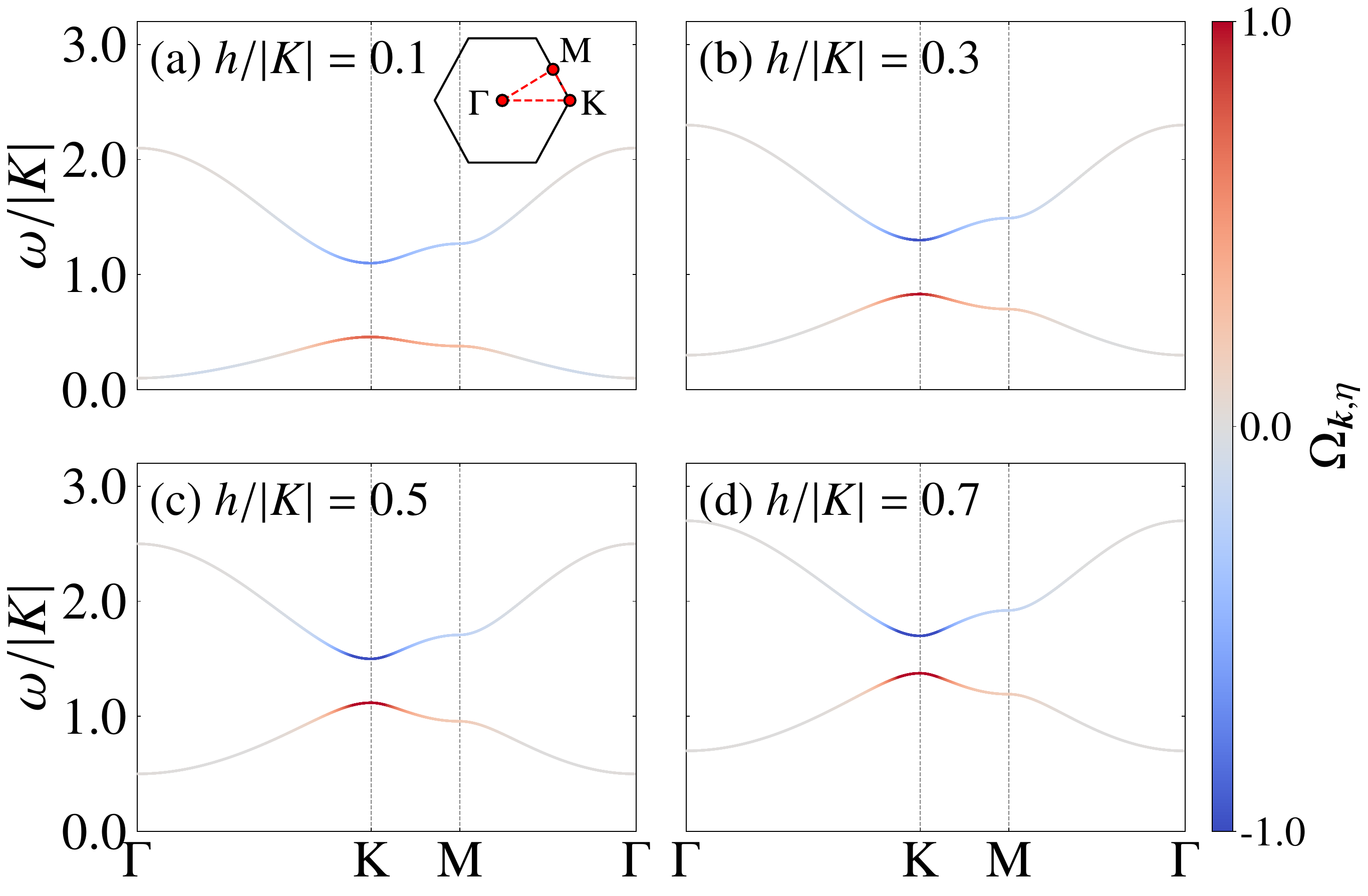}
    \caption{
    Dispersion relations of magnons from a spin polarized state in the Kitaev model with (a) $h/|K|=0.1$, (b) $0.3$, (c) $0.5$, and (d) $0.7$.
The line color indicates the Berry curvature $\Omega_{\bm{k},\eta}$.
The inset in (a) shows the first Brillouin zone of the honeycomb lattice.
The dispersion relations are plotted along the red dashed lines in this inset.
    }
    \label{fig:bc_kitaev}
  \end{center}
\end{figure}

Before showing the results for the thermal Hall conductivity, we briefly comment on the magnon band structure under several magnetic fields.
Figures~\ref{fig:bc_kitaev}(a)--\ref{fig:bc_kitaev}(d) present the dispersion relations of magnons for $h/|K|=0.1$, $0.3$, $0.5$, and $0.7$, respectively~\cite{mcclarty2018}.
There are two magnon branches in this system, and our analysis has confirmed that the Chern numbers of low-energy and high-energy branches are $+1$ and $-1$, respectively. 
Across all parameters, the absolute value of the Berry curvature around the K point takes a large value.
We observe that with increasing the magnetic field, the magnon dispersion shifts to the high-energy side and the gap between the two bands becomes narrow.
Despite the Chern number of the low-energy branch being $+1$, the Berry curvature at this branch around the $\Gamma$ point takes a small negative value, particularly for $h/|K|=0.1$~\cite{mcclarty2018}.

\begin{figure}[t]
  \begin{center}
    \includegraphics[width=\columnwidth,clip]{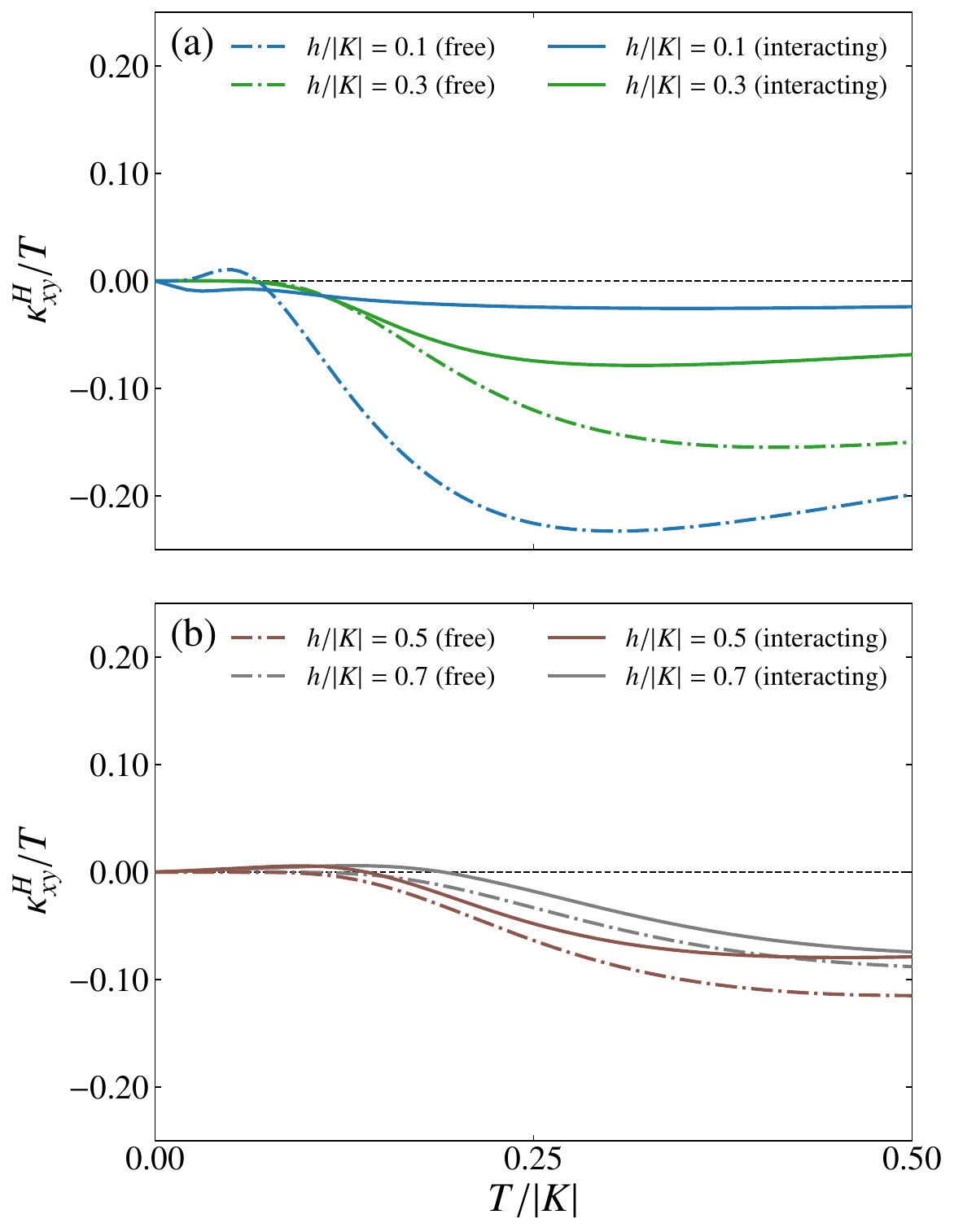}
    \caption{
    Temperature dependence of thermal Hall conductivity divided by temperature in the Kitaev model for (a) $h/|K|=0.1$ and $0.3$ and (b) $h/|K|=0.5$ and $0.7$.
The dashed-dotted lines represent the results for the free magnon system within the linear spin-wave approximation.
On the other hand, the solid lines represent the results for the systems with magnon-magnon interactions calculated based on the iDE approach.
    }
    \label{fig:kxy_kitaev}
  \end{center}
\end{figure}

Here, we present the temperature dependence of the thermal Hall conductivity $\kappa_{xy}^{H}$ calculated with the magnon damping in Fig.~\ref{fig:kxy_kitaev}.
For comparison, we also provide results obtained under the free-magnon approximation based on Eq.~\eqref{eq:kxy-free}.
First, we review the results in the free-magnon system~\cite{mcclarty2018}.
The thermal Hall conductivity takes a negative value in the temperature range except for extremely low temperatures, and $\kappa_{xy}^{H}/T$ asymptotically approaches zero at high temperatures.
This phenomenon can be attributed to the overall structure of Berry curvature associated with magnon bands and functional form of $c_{2}(g(\beta\varepsilon_{\bm{k},\eta}))$; the low-energy band with the positive Chern number largely contributes to the thermal Hall effect compared to the high-energy band because $c_{2}(g(\beta\varepsilon_{\bm{k},\eta}))$ rapidly decreases with increasing $\varepsilon_{\bm{k},\eta}$ at low temperatures.
This feature results in the negative value of $\kappa_{xy}^{H}$ because of the negative sign in Eq.~\eqref{eq:kxy-free}.
At high temperatures, $c_{2}(g(\beta\varepsilon_{\bm{k},\eta}))$ is almost independent on $\varepsilon_{\bm{k},\eta}$, and thereby, contributions from two bands with opposite Chern numbers to the thermal Hall conductivity chancel out each other.
On the other hand, in the low-temperature region, the thermal Hall conductivity turns to be positive at $h/|K|=0.1$, as shown in Fig.~\ref{fig:kxy_kitaev}(a)~\cite{mcclarty2018}.
As mentioned before, the sign change of $\kappa_{xy}^{H}$ is ascribed to the negative Berry curvature of the low-energy branch in the vicinity of the $\Gamma$ point.
We also find that the absolute value of $\kappa_{xy}^{H}/T$ decreases with increasing the external magnetic field, as seen in Fig.~\ref{fig:kxy_kitaev}(b).
This trend can be comprehended through the increase of the excitation energy and the narrowing of the gap energy between the two branches in Fig.~\ref{fig:bc_kitaev}.

\begin{figure*}[t]
  \begin{center}
    \includegraphics[width=2\columnwidth,clip]{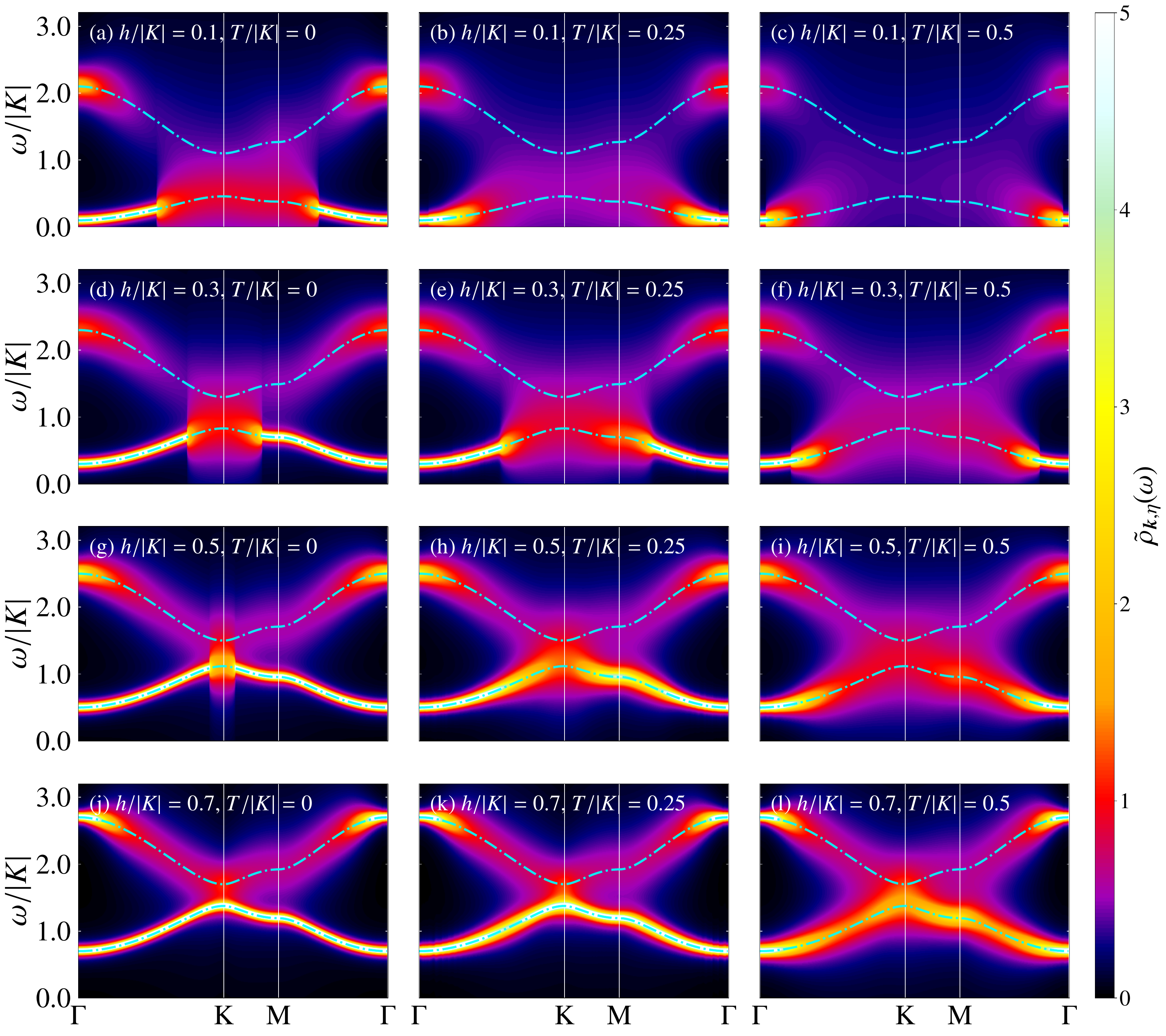}
    \caption{
    (a)--(c) Color map of the spectral function $\tilde{\rho}_{\bm{k},\eta}(\omega)$ calculated by the iDE approach in the Kitaev model at (a) $T/|K|=0$, (b) $0.25$, (c) $0.5$ for $h/|K|=0.1$.
    The cyan dashed-dotted lines stand for the dispersion relations obtained by the linear spin-wave approximation.
(d)--(f), (g)--(i), (j)--(l) correspond to the results at $h/|K|=0.3$, $0.5$, and $0.7$, respectively.
The spectral functions are plotted along the red dashed lines in the inset of Fig.~\ref{fig:bc_kitaev}(a).
    }
    \label{fig:spec_kitaev}
  \end{center}
\end{figure*}

Next, we discuss the temperature dependence of the thermal Hall conductivity incorporating magnon damping.
We find that the thermal Hall conductivity is strongly suppressed by the magnon damping, especially in the systems under low magnetic fields.
The sign change in $\kappa^H_{xy}$, predicted by calculations using the free-magnon approximation, vanishes at low temperatures when $h/|K|=0.1$.
This is understood from the approximated spectral function $\tilde{\rho}_{\bm{k},\eta}(\omega)$.
From Eq.~\eqref{eq:kxy-L}, the thermal Hall conductivity depends on this quantity with the damping rate $\tilde{\Gamma}_{\bm{k},\eta}$ in addition to the Berry curvature.
Figures~\ref{fig:spec_kitaev}(a)--\ref{fig:spec_kitaev}(c) show the spectral function $\tilde{\rho}_{\bm{k},\eta}(\omega)$ on the $\bm{k}$-$\omega$ plane for $h/|K|=0.1$.
At $T=0$, the lower-energy mode survives around the $\Gamma$ point, which is responsible for the sign change of the thermal Hall conductivity in the free-magnon approximation.
On the other hand, the two magnon modes near the $\mathrm{K}$-$\mathrm{M}$ path are strongly damped.
Note that the Berry curvature in the lower-energy mode around the $\mathrm{K}$ point takes a large positive value in the free-magnon approximation [Fig.~\ref{fig:bc_kitaev}(a)].
The broadened spectrum contributes to the spectral weight near zero energy, which results in a negative thermal Hall conductivity.
Therefore, the thermal Hall conductivity remains negative even at low temperatures.
With increasing temperature, the smearing becomes more pronounced, which causes the strong suppression of $\kappa^H_{xy}/T$ at higher temperatures, as discussed in Sec.~\ref{sec:thermal-Hall-damping}.

As the external magnetic field increases, the difference between $\kappa^H_{xy}$ with and without magnon-magnon interactions diminishes.
This behavior arises from the reduction of magnon damping due to the application of the magnetic field, as depicted in Fig.~\ref{fig:spec_kitaev}.
The magnon damping at low temperatures is primarily attributed to a decay process in which a magnon splits into two magnons~\cite{koyama2023}.
Applying the magnetic field decreases the overlap between the magnon branch and the two-magnon continuum, thereby suppressing magnon-magnon scatterings.
Similar phenomena have been observed in the damping of a chiral magnon edge mode in the Kitaev model on a ribbon-shaped cluster, where the damping rate decreases monotonically with increasing a magnetic field~\cite{koyama2023}.
These results suggest a close relationship between the decay of the chiral edge mode and the impact of magnon-magnon interactions on thermal Hall conductivity.

We also find that a sign change occurs when $h/|K|=0.5$ and $0.7$, as shown in Fig.~\ref{fig:kxy_kitaev}(b).
This phenomenon can be attributed the difference between the magnon damping for the two branches.
As shown in Fig.~\ref{fig:spec_kitaev}(j), even at zero temperature, magnons within the high-energy branch undergo significant damping, primarily due to the magnon decay process split into two magnons~\cite{koyama2023}.
As discussed in Sec.~\ref{sec:thermal-Hall-damping}, pronounced magnon damping amplifies the impact of the Berry curvature in the corresponding branch on thermal Hall conductivity.
Consequently, the sign change to a positive value of $\kappa^H_{xy}$ at low temperatures, depicted in Fig.\ref{fig:kxy_kitaev}(b), results from the magnon damping in the high-energy branch with negative Berry curvatures, as demonstrated in Fig.\ref{fig:bc_kitaev}(d).

\subsection{Heisenberg-DM model}
\label{sec:Heisenberg-DM}

\begin{figure}[t]
  \begin{center}
    \includegraphics[width=\columnwidth,clip]{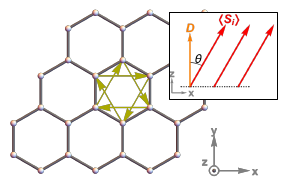}
    \caption{
    Schematic picture of the honeycomb lattice on which the Heisenberg-DM model is defined.
    The yellow arrows represent the vectors connecting next-nearest sites clockwise.
    The inset shows the definition of $\theta$, which is an angle between the direction of the DM vector and ferromagnetic spin moment.
    }
    \label{fig:honeycomb_DMI}
  \end{center}
\end{figure}

In this section, we examine the thermal Hall response in an $S=1/2$ quantum spin model with Heisenberg and DM interactions on a honeycomb lattice, which is represented as~\cite{habel2024},
\begin{align}
    \label{eq:heisen-dm-model}
    \mathcal{H} = J \sum_{\braket{ij}} \bm{S}_{i}\cdot \bm{S}_{j} + \sum_{\braket{\braket{ij}}}\bm{D}_{ij} \cdot  \left( \bm{S}_{i}\times \bm{S}_{j} \right),
\end{align}
where $J$ and $\bm{D}_{ij}$ are the exchange constant of the Heisenberg interaction between nearest-neighbor sites $\braket{ij}$ and the DM vector for the bond $\braket{\braket{ij}}$ connecting next nearest-neighbor sites $i$ and $j$, respectively.
In this model, an $S=1/2$ spin at site $i$ is represented by $\bm{S}_i=(S_i^x,S_i^y,S_i^z)$.
Unlike the Kitaev model, the axes of the spin space are aligned with those in real space (see Fig.~\ref{fig:honeycomb_DMI}).
We assume that the Heisenberg interaction is ferromagnetic ($J<0$), and the DM vector is parallel to the $z$ axis, $\bm{D}_{ij} = (0,0,\pm D)$, where the plus (minus) sign is assigned when the vector connecting from site $i$ to $j$ is oriented clockwise (anticlockwise) in a hexagon plaquette of the honeycomb lattice (see Fig.~\ref{fig:honeycomb_DMI}).
In the present calculations, we choose $D=0.3|J|$.
In this case, the classical ground state is a fully polarized ferromagnetic state and is degenerate for the direction of the spin polarization~\cite{habel2024}.
Here, we introduce the parameter $\theta$ to denote a tilting angle of the spin moment from the $z$ axis to the $x$ axis (see the inset of Fig.~\ref{fig:honeycomb_DMI}).
Note that, at $\theta = 0$, the total spin operator $\sum_{i} S_{i}^{z}$ commutes with the Hamiltonian, preventing the emergence of magnon-magnon interactions without particle-number conservation in the nonlinear spin-wave theory~\cite{chernyshev2016}.

\begin{figure}[t]
  \begin{center}
    \includegraphics[width=\columnwidth,clip]{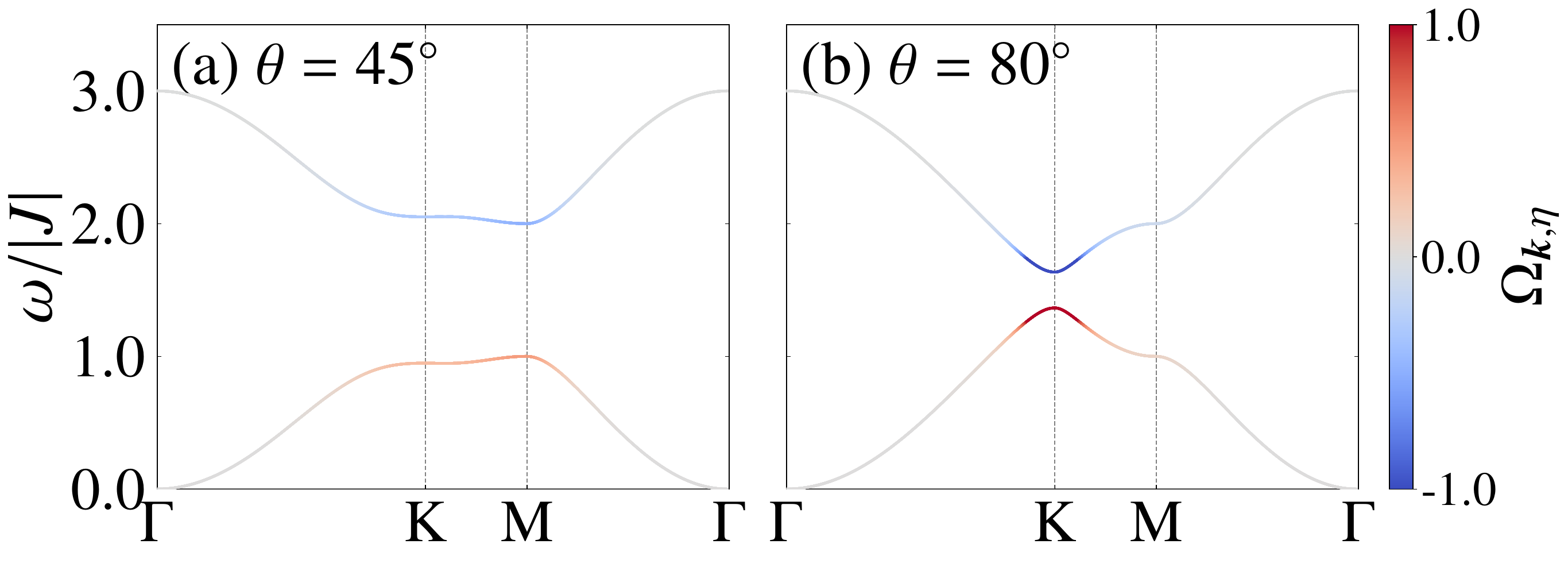}
    \caption{
    Dispersion relations of magnons in the Heisenberg-DM model from a ferromagnetic state
    with (a) $\theta=45^{\circ}$ and (b) $80^{\circ}$.
The line color indicates the Berry curvature $\Omega_{\bm{k},\eta}$.
The dispersion relations are plotted along the red dashed lines in the inset of Fig.~\ref{fig:bc_kitaev}(a).
    }
    \label{fig:bc_DMI}
  \end{center}
\end{figure}

Figures~\ref{fig:bc_DMI}(a) and \ref{fig:bc_DMI}(b) illustrate the dispersion relations of magnons at $\theta=45^\circ$ and $\theta=80^\circ$, respectively.
In both cases, the dispersion relations exhibit two branches, with the absolute value of the Berry curvature increasing significantly near the K-M path, while approaching zero in the vicinity of the $\Gamma$ point.
We have confirmed that the Chern numbers of low-energy and high-energy branches are $+1$ and $-1$, respectively.
We also find that the gap between the two magnon bands narrows with increasing $\theta$.

\begin{figure}[t]
  \begin{center}
    \includegraphics[width=\columnwidth,clip]{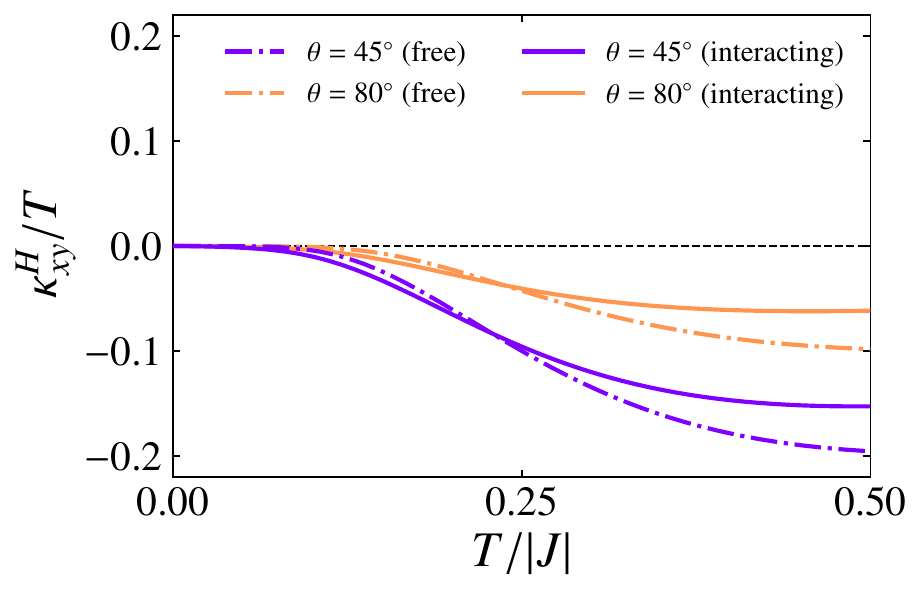}
    \caption{
    Temperature dependence of thermal Hall conductivity divided by temperature in the Heisenberg-DM model from a ferromagnetic state with $\theta=45^{\circ}$ and $80^{\circ}$.
The dashed-dotted lines represent the results for the free magnon system within the linear spin-wave approximation.
On the other hand, the solid lines represent the results for the systems with magnon-magnon interactions calculated based on the iDE approach.
    }
    \label{fig:kxy_DMI}
  \end{center}
\end{figure}

Figure~\ref{fig:kxy_DMI} displays the temperature dependence of $\kappa^H_{xy}/T$ both with and without magnon-magnon interactions for $\theta=45^\circ$ and $80^\circ$.
Given that the Chern number for the lower-energy branch is positive, $\kappa^H_{xy}$ exhibits negative values across a broad temperature range.
We find that the absolute value of $\kappa^H_{xy}$ for $\theta=80^\circ$ is smaller than that for $\theta=45^\circ$.
This phenomenon is understood from the reduction of the magnon gap between the two bands with increasing $\theta$.
As discussed in Sec.~\ref{sec:thermal-Hall-damping}, decreasing the magnon gap leads to suppressing the thermal Hall conductivity.

\begin{figure*}[t]
  \begin{center}
    \includegraphics[width=170mm,clip]{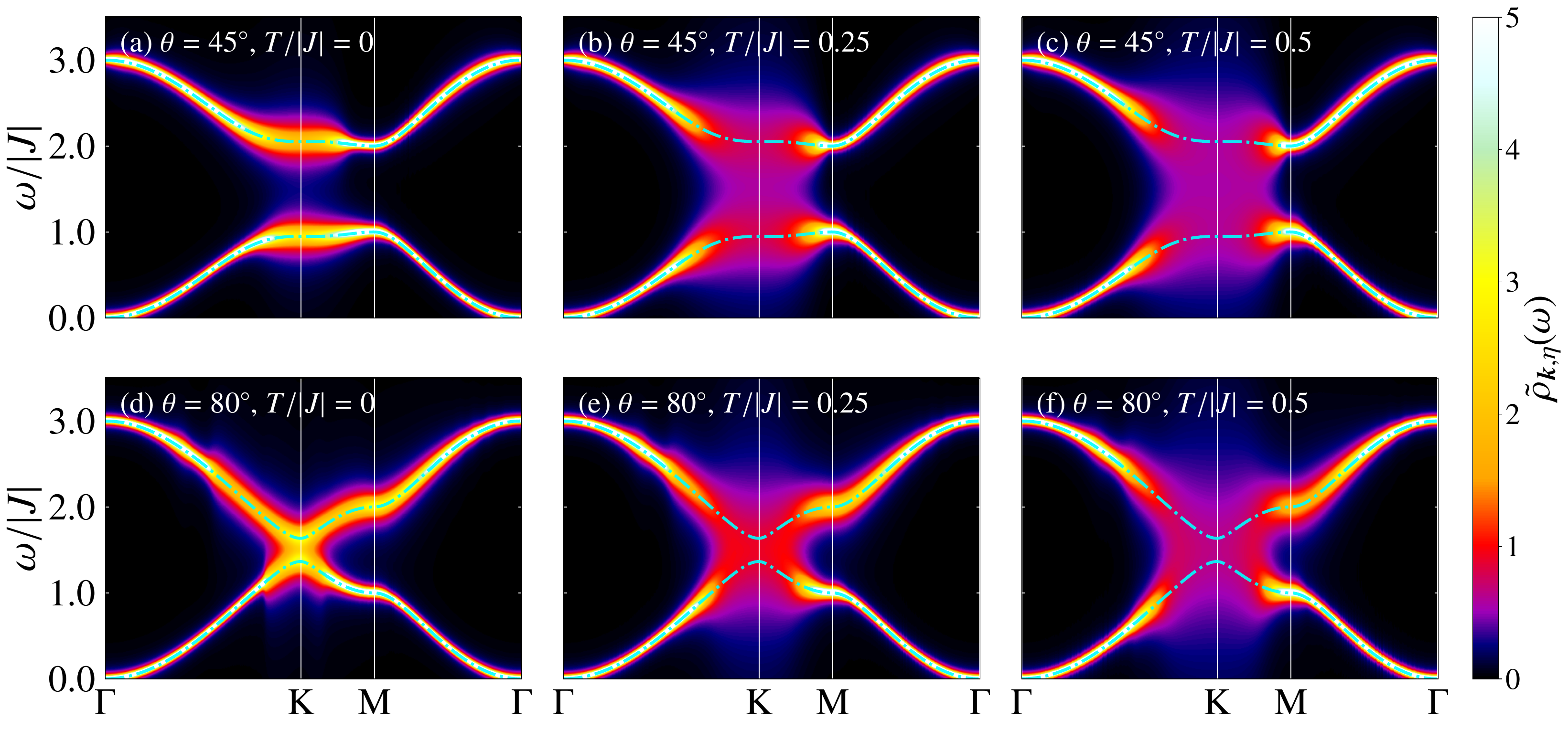}
    \caption{
    (a)--(c) Color map of the spectral function $\tilde{\rho}_{\bm{k},\eta}(\omega)$ calculated by the iDE approach in the Heisenberg-DM model at (a) $T/|J|=0$, (b) $T/|J|=0.25$, (c) $T/|J|=0.5$ for $\theta=45^{\circ}$.
    The cyan dashed-dotted lines stand for the dispersion relations obtained by the linear spin-wave approximation.
(d)--(f) Corresponding results for $\theta=80^{\circ}$.
The spectral functions are plotted along the red dashed lines in the inset of Fig.~\ref{fig:bc_kitaev}(a).
    }
    \label{fig:spec_DMI}
  \end{center}
\end{figure*}

Next, we discuss the effect of the magnon-magnon interaction on the thermal Hall conductivity.
As shown in Fig.~\ref{fig:kxy_DMI}, this effect slightly enhances the absolute value of $\kappa^H_{xy}$ at low temperatures but suppresses it at higher temperatures.
This behavior is expected from the simplified model with magnon damping introduced in Sec.~\ref{sec:thermal-Hall-damping}.
To examine the contribution of magnon-magnon interactions, we calculate the spectral function of magnons calculated by the iDE approach.
The color map of the  spectral function is presented in Fig.~\ref{fig:spec_DMI} at several temperatures.
In both cases with $\theta=45^\circ$ and $80^\circ$, magnon damping occurs around the K point.
Given that magnon bands near this point possess large values of the Berry curvature, it is found that magnon damping significantly influences the thermal Hall conductivity. 
It has been demonstrated that a chiral edge mode is strongly damped by incorporating magnon-magnon interactions in a cluster with open boundaries~\cite{habel2024}.
Our findings suggest that the damping of the chiral edge mode corresponds to the significant contribution of magnon-magnon interactions to the thermal Hall conductivity found in the present study, which is similar to the case of the Kitaev model introduced in the previous section.

\section{Summary and Discussion}
\label{summary}
In summary, we have derived the expression for the thermal Hall conductivity incorporating the magnon damping in localized electron systems based on nonlinear spin-wave theory, where magnons are introduced as elementary excitations from a magnetic order.
We have formulated the thermal response by accounting for both the Kubo formula and heat magnetization based on Green's functions of magnons.
The effect of the magnon damping is introduced as the imaginary part of the self-energy, which gives rise to the broadening of the magnon spectrum.
The thermal Hall conductivity obtained in the present study reproduces the previous result of free-magnon systems in the zero limit of the magnon damping.
Based on the expression of the thermal Hall conductivity, we first discussed the impact of magnon damping in a simple magnon model with nonzero Chern numbers.
We have found that the magnon damping slightly enhances the thermal Hall conductivity at very low temperatures due to the increase of the low-energy spectral weight of magnons resulting from the spectrum broadening.
Meanwhile, the thermal Hall conductivity is suppressed by the magnon damping at higher temperatures by the cancellation of contributions from higher-energy magnon branches with Berry curvatures taking opposite values, which is also caused by the broadening of the magnon spectrum.
We have also applied the present theory to two localized spin models on a honeycomb lattice: the Kitaev model under magnetic fields and the ferromagnetic Heisenberg model with Dzyaloshinskii-Moriya interactions.
In these models, the imaginary part of the self-energy, which arises from the magnon-magnon interactions beyond the linear spin-wave theory, has been evaluated by the self-consistent imaginary Dyson equation approach at finite temperatures.
We have clarified that magnon damping substantially affects the temperature dependence of the thermal Hall conductivity in both systems.
In particular, in the presence of significant quantum fluctuations, the low-energy magnon branches largely decay, and the absolute value of the thermal Hall conductivity is strongly reduced from the value obtained in the free-magnon system.
We have found that such a substantial change of the thermal Hall conductivity occurs when a chiral edge mode is largely damped,  suggesting the presence of bulk-edge correspondence even in the presence of magnon-magnon interactions.

Since our study begins with a general form of localized electron systems with multiple local degrees of freedom, the present results are easily applied to other models, such as localized systems with multipole interactions including spin-orbital systems, spin dimer systems typified by the Shastry-Sutherland model, skyrmion crystals, and localized electron systems coupled with lattice vibrations.
In this study, we have focused on elucidating the impact of magnon damping by considering only the imaginary part of the self-energy on the thermal Hall conductivity.
On the other hand, this study has yet to incorporate the real part of the self-energy, which shifts the magnon energy, and the effects of vertex correction.
These contributions will be addressed in future work.
Additionally, formulating other thermal responses, exemplified by the spin Nernst effect, remains challenging for future research.

\begin{acknowledgments}
The authors thank S.~Murakami, M.~Zhitomirsky, A.~Shitade, and A.~Ono for fruitful discussions. 
Parts of the numerical calculations were performed in the supercomputing
systems in ISSP, the University of Tokyo.
This work was supported by Grant-in-Aid for Scientific Research from
JSPS, KAKENHI Grant No.~JP19K03742, JP20H00122, JP22H01175, JP23H01129, and JP23H04865,
and by JST, the establishment of university fellowships towards the creation of science technology innovation, Grant Number JPMJFS2102.
\end{acknowledgments}

\appendix
\section{Definition of heat magnetization}
\label{app:def-orbmag}

When we calculate a response for a statistical force (e.g., thermal gradient) applied to a system, we must pay attention to the impact of such a gradient on the rotational motion intrinsic to the wave packet of carriers~\cite{xiao2006}.
In the present paper, we focus on magnon systems with zero chemical potential, and hence, we omit the effect of the chemical-potential gradient and focus on the thermal transport.
In the thermal Hall effect, the contribution of heat magnetization originating from the rotational motion appears in addition to that evaluated from the Kubo formula~\cite{qin2011}.
When the system is in equilibrium without thermal gradients, the thermal current density satisfies $\nabla \cdot \left\langle \bm{j}^{Q} (\bm{r}) \right\rangle = 0$.
In this case, the rotational motion of carriers only contributes to the thermal current density.
Thus, we define the heat magnetization density $\bm{m}^{Q} (\bm{r})$ by~\cite{qin2011},
\begin{align}
    \left\langle \bm{j}^{Q}(\bm{r}) \right\rangle = \nabla \times \bm{m}^{Q} (\bm{r}).
    \label{appeq:def-mQ}
\end{align}
In a similar manner, the macroscopic transport thremal current $\bm{J}^{\mathrm{tr}}$ in the presence of the gravitational-field gradient $\nabla\chi$ is introduced as
\begin{align}
    \bm{J}^{\mathrm{tr}} = \frac{1}{V}\int d\bm{r}
    \left[ \left\langle \bm{j}^{Q;\chi}(\bm{r}) \right\rangle_{\nabla \chi} - \nabla\times \bm{m}^{Q;\chi}(\bm{r}) \right],
\end{align}
where $\braket{\ \cdot \ }_{\nabla \chi}$ and $\bm{m}^{Q;\chi}(\bm{r})$ represent the expectation value and heat magnetization density in the presence of $\nabla\chi$.
Up to the first order of $\nabla \chi$, the first term of the above equation is written as
\begin{align}
    \frac{1}{V}\int d\bm{r}\left\langle \bm{j}^{Q;\chi}(\bm{r}) \right\rangle_{\nabla \chi}\simeq
    \left\langle \bm{J}^{Q} \right\rangle_{\nabla \chi}+\frac{1}{V}\int d\bm{r}\left\langle \bm{j}^{Q;\chi}(\bm{r}) \right\rangle,
\end{align}
where $\bm{J}^{Q}=\frac{1}{V}\int d\bm{r}\bm{j}^{Q}(\bm{r})$.
The first term can be evaluated by the Kubo formula as
\begin{align}
    \left\langle \bm{J}^{Q} \right\rangle_{\nabla \chi}\simeq \sum_{\lambda'}S_{\lambda\lambda'}(-\nabla_{\lambda'} \chi).
\end{align}
On the other hand, $\left\langle \bm{j}^{Q;\chi}(\bm{r}) \right\rangle$ in the second term is calculated as
\begin{align}
    \left\langle \bm{j}^{Q;\chi}(\bm{r}) \right\rangle\simeq \nabla\times \bm{m}^{Q;\chi}(\bm{r})-2\bm{m}^{Q}(\bm{r})\times\nabla \chi,
\end{align}
up to the first order of $\chi$.
From these expressions, we obtain Eqs.~\eqref{eq:thermal coefficient-chi} and \eqref{eq:Lxy}.
Here, the total heat magnetization is introduced as $\bm{M}^{Q} = \int \bm{m}^{Q}(\bm{r})d\bm{r}$.
This quantity is evaluated from Eqs.~\eqref{eq:def-Mz} and \eqref{eq:def-tildeMz}, which are derived from Eq.~\eqref{appeq:def-mQ}~\cite{qin2011}.

\section{Thermal current operator and scaling low}
\label{app:tco and sl}
\subsection{Thermal current operator}

In this section, we derive the representation of the thermal current density operator given in Eq.~\eqref{eq:scaling-jq} from the bilinear bosonic Hamiltonian ${\cal H}_0$ in Eq.~\eqref{eq:Hamil-A-real}.
We start from the following equation of continuity in a continuum limit:
\begin{align}
    \label{appeq:energy-continuity}
    \pdiff{h^{\chi}(\bm{r})}{t} = -\frac{i}{\hbar} \left[ h^{\chi} (\bm{r}), \mathcal{H}^{\chi} \right]
    = - \nabla\cdot \bm{j}^{Q;\chi} (\bm{r}).
\end{align}
where $h^{\chi}(\bm{r})$, $\mathcal{H}^{\chi}$, and $\bm{j}^{Q;\chi} (\bm{r})$ are the local Hamiltonian, total Hamiltonian, and thermal current density in the presence of the gravitational field $\chi (\bm{r})$, which are defined as follows.
In the continuum limit, the bosonic Hamiltonian ${\cal H}_0$ without the gravitational field is represented as~\cite{matsumoto2014}
\begin{align}
    \mathcal{H}_{0} &= \frac{1}{2} \sum_{\bm{\delta}} \int d\bm{r} \mathcal{A}^\dagger (\bm{r}) \mathcal{M}_{\bm{\delta}} \mathcal{A}^{}(\bm{r}+\bm{\delta})\nonumber\\
    &= \frac{1}{2} \int d\bm{r} \mathcal{A}^\dagger (\bm{r}) \hat{\mathcal{M}}_{0} \mathcal{A}^{}(\bm{r}) \label{appeq:H0}
\end{align}
where $\hat{\mathcal{M}}_{0}$ is given by
\begin{align}
    \hat{\mathcal{M}}_{0} = \sum_{\bm{\delta}}
        \mathcal{M}_{\bm{\delta}} e^{i\hat{\bm{p}}\cdot \bm{\delta}/\hbar},
        \label{appeq:Mdelta}
\end{align}
Here, $\mathcal{M}_{\bm{\delta}}$ is a $2N\times 2N$ Hermitian matrix depending on $\bm{\delta}$ 
owing to the translational symmetry and represented as
\begin{align}
    \mathcal{M}_{\bm{\delta}} =
    \begin{pmatrix}
        \mathcal{M}_{\bm{\delta}}^{11} &\ \mathcal{M}_{\bm{\delta}}^{12}\\
        \left(\mathcal{M}_{\bm{\delta}}^{12}\right)^{*} &\ \left(\mathcal{M}_{-\bm{\delta}}^{11}\right)^{T}
    \end{pmatrix}.
\end{align}
This corresponds to Eq.~\eqref{eq:Mdelta} for lattice systems.
Since $\hat{\mathcal{M}}_{0}$ is a Hermitian matrix, $\mathcal{M}_{\bm{\delta}}$ 
satisfies the relation $\mathcal{M}_{\bm{\delta}}^{\dagger} = \mathcal{M}_{-\bm{\delta}}$.
We also introduce $\mathcal{A} (\bm{r})$ as a set of the $2N$ bosonic operators, which is given by
\begin{align}
    \mathcal{A}_{s} (\bm{r})
    =
    \begin{cases}
            a_{s} (\bm{r}) & (s = 1,\cdots , N)\\
    a_{s-N}^\dagger (\bm{r}) & (s = N+1,\cdots , 2N)
    \end{cases}
\end{align}
where $a_{s}(\bm{r})$ and $a_{s}^\dagger(\bm{r})$ are annihilation and creation operators satisfying the commutation relations such as $\left[ a_{s}^\dagger (\bm{r}), a_{s'}^{}(\bm{r}') \right] = \delta_{ss'} \delta(\bm{r}-\bm{r}')$.
Moreover, the operator $\hat{\bm{p}}$ in Eq.~\eqref{appeq:Mdelta} is defined as a generator of translation for the bosonic operator $\mathcal{A}_{s} (\bm{r})$, which satisfies the relation $e^{i\hat{\bm{p}}\cdot\bm{\delta}/\hbar} \mathcal{A}_{s}(\bm{r}) = \mathcal{A}_{s} (\bm{r}+\bm{\delta})$.
From Eq.~\eqref{appeq:H0}, the local Hamiltonian for $\mathcal{H}_0$ can be written as 
\begin{align}
    h(\bm{r}) = \frac{1}{2}\mathcal{A}^\dagger (\bm{r}) \hat{\mathcal{M}}_{0} \mathcal{A}^{} (\bm{r}).
\end{align}
In a similar manner, the local Hamiltonian $h^{\chi} (\bm{r})$ in the presence of the gravitational field, which satisfies $\mathcal{H}^{\chi}= \int d\bm{r}h^{\chi} (\bm{r})$, is represented as,
\begin{align}
    h^{\chi} (\bm{r})&=  h(\bm{r}) + \frac12[\chi(\bm{r}) h(\bm{r}) + h(\bm{r})\chi(\bm{r})]\notag\\
    &\simeq \frac{1}{2}\tilde{\mathcal{A}}^\dagger(\bm{r})\hat{\mathcal{M}}_{0}\tilde{\mathcal{A}}^{} (\bm{r}),
    \label{appeq:h_chi}
\end{align}
where $\tilde{\mathcal{A}}^{} (\bm{r}) =  \left[1+\frac{\chi(\bm{r})}{2}\right] \mathcal{A}^{}(\bm{r})$.
Note that we apply symmetrization for $\chi(\bm{r})$ and $h(\bm{r})$ because these do not commute due to the operator $\hat{\bm{p}}$ in $h(\bm{r})$ in the continuum limit~\cite{matsumoto2014}.
Using the local Hamiltonian, we introduce the energy polarization $\bm{P}^{\chi}_{E}$ as
\begin{align}
    \bm{P}^{\chi}_{E} = \frac{1}{2}\int d\bm{r} \left[ \bm{r}h^{\chi}(\bm{r}) + h^{\chi}(\bm{r})\bm{r} \right].
\end{align}
From Eq.~\eqref{eq:def-J}, we calculate the thermal current as follows:
\begin{widetext}
\begin{align}
    \bm{J}^{Q;\chi} &= \frac{i}{V\hbar} \left[\mathcal{H}^{\chi}, \bm{P}_{E} \right]
    = \frac{1}{4V} \int d\bm{r} 
    \tilde{\mathcal{A}}^{\dagger}(\bm{r}) 
    \left\{
    \hat{\bm{v}} \sigma_3 \left[1 + \frac{\chi(\bm{r})}{2}\right]^2 \hat{\mathcal{M}}_{0}
    +    
    \hat{\mathcal{M}}_{0} \left[1 + \frac{\chi(\bm{r})}{2}\right]^2 \sigma_3 \hat{\bm{v}} 
    \right\}
    \tilde{\mathcal{A}}^{} (\bm{r}),
\end{align}
where we use the following relations $\left(\mathcal{M}_{\bm{\delta}}\right)_{m,n} = \left(\mathcal{M}_{-\bm{\delta}}\right)_{n+N,m+N}$ and $\left(\mathcal{M}_{\bm{\delta}}\right)_{m,n+N} = \left(\mathcal{M}_{-\bm{\delta}}\right)_{m+N,n}$, corresponding to Eqs.~\eqref{eq:M11} and \eqref{eq:M12}, and $\mathcal{A}_{m}^{}(\bm{r}) = \mathcal{A}_{m+N}^\dagger (\bm{r})$ and $\mathcal{A}_{m}^\dagger(\bm{r}) = \mathcal{A}_{m+N}^{} (\bm{r})$ for $n,m = 1,2,\cdots,N$.
The velocity $\hat{\bm{v}}$ is given by
\begin{align}
    \hat{\bm{v}} = - \frac{i}{\hbar}\left[ \bm{r},\hat{\mathcal{M}}_{0} \right] = \frac{i}{\hbar} \sum_{\bm{\delta}} \bm{\delta} \mathcal{M}_{\bm{\delta}} 
    e^{i\hat{\bm{p}}\cdot\bm{\delta}/\hbar}.
\end{align}
Thus, the thermal current densities in the absence and presence of $\chi$ are expressed by
\begin{align}
    \label{appeq:jQ}
    \bm{j}^{Q} (\bm{r}) &= \frac{1}{4} \mathcal{A}^\dagger (\bm{r}) 
    \left(
    \hat{\bm{v}} \sigma_3  \hat{\mathcal{M}}_{0}
    +
    \hat{\mathcal{M}}_{0}  \sigma_3 \hat{\bm{v}}
    \right)
    \mathcal{A}^{}(\bm{r}),\\
    \label{appeq:jQchi}
    \bm{j}^{Q;\chi} (\bm{r}) &= \frac{1}{4} \mathcal{A}^\dagger (\bm{r}) 
    \left[ 1 + \frac{\chi(\bm{r})}{2} \right]
    \left\{
    \hat{\bm{v}} \sigma_3 \left[ 1 + \frac{\chi(\bm{r})}{2} \right]^2 \hat{\mathcal{M}}_{0}
    +
    \hat{\mathcal{M}}_{0} \left[ 1 + \frac{\chi(\bm{r})}{2} \right]^2 \sigma_3 \hat{\bm{v}}
    \right\}
    \left[ 1 + \frac{\chi(\bm{r})}{2} \right]
    \mathcal{A}^{}(\bm{r}),
\end{align}
respectively.
Using the relation $\chi(\bm{r}) = \bm{r} \cdot \nabla \chi(\bm{r})$ with $\nabla \chi$ being constant, we expand Eq.~\eqref{appeq:jQchi} with respect to $\nabla \chi$ as
\begin{align}
    \bm{j}^{Q;\chi} (\bm{r}) = \bm{j}^{Q}(\bm{r}) + \bm{j}^{Q}_{\nabla\chi}(\bm{r}).
\end{align}
Here, $\bm{j}^{Q}_{\nabla\chi}(\bm{r})$ is the term proportional to $\nabla \chi$ in $\bm{j}^{Q;\chi}$, which is represented as
\begin{align}
    \bm{j}^{Q}_{\nabla\chi}(\bm{r})
    = &-\frac{i\hbar}{8} \sum_\lambda\left(\nabla_{\lambda} \chi \right)\mathcal{A}^\dagger (\bm{r}) 
    \left( \hat{\bm{v}} \sigma_3 \hat{v}_{\lambda} -  \hat{v}_{\lambda} \sigma_3 \hat{\bm{v}} \right)
    \mathcal{A}^{} (\bm{r})\nonumber\\
    \label{appeq:jQnablachi}
    &+ 
    \frac{1}{8}\sum_\lambda\left(\nabla_{\lambda} \chi \right)
    \left[ 
    \mathcal{A}^\dagger (\bm{r}) \left( r_{\lambda} \hat{\bm{v}} \sigma_3 + 3\hat{\bm{v}}\sigma_3 r_{\lambda} \right) \hat{\mathcal{M}}_{0} \mathcal{A}^{}(\bm{r})
    + \mathcal{A}^\dagger (\bm{r}) \hat{\mathcal{M}}_{0} \left( 3r_{\lambda} \sigma_3 \hat{\bm{v}} + \sigma_3 \hat{\bm{v}} r_{\lambda} \right) \mathcal{A}(\bm{r})
    \right].
\end{align}

\subsection{Scaling law for thermal current operator}

To evaluate the heat magnetization $M^{Q}_{z}$, we impose the scaling relation given in Eq.~\eqref{eq:scaling-j} for the current density operator.
Similar to Eq.~\eqref{appeq:h_chi}, the scaling relation is applied as the following symmetric form in the continuum limit:
\begin{align}
    \bm{j}^{Q;\chi}(\bm{r}) = 
    \frac{1}{2}
    \left\{
    \left[1 + \chi(\bm{r}) \right]^2 \bm{j}^{Q} (\bm{r}) + \bm{j}^{Q}(\bm{r}) \left[1 + \chi(\bm{r}) \right]^2 
    \right\}\label{appeq:scaling}
\end{align}
From Eq.~\eqref{appeq:energy-continuity}, the equation of continuity is invariant under the gauge transformation inherent in the thermal current density as follows:
\begin{align}
    \bm{j}^{Q} \to \bm{j}^{Q} + \nabla \times \bm{f}(\bm{r}),
\end{align}
where $\bm{f}(\bm{r})$ is an arbitrary vector function.
Using the redundant degrees of freedom, we determine the expression of thermal current density so as to satisfy the scaling relation given in Eq.~\eqref{appeq:scaling}.
Here, by applying $\bm{r}\cdot \nabla\chi = \chi(\bm{r})$ to the second term of Eq.~\eqref{appeq:jQnablachi}, we can rewrite $\bm{j}^{Q;\chi}(\bm{r})$ up to the first order of $\chi$ as
\begin{align}
    \bm{j}^{Q;\chi} (\bm{r}) = &
    \frac{\left[1 + \chi(\bm{r}) \right]^2 \bm{j}^{Q} (\bm{r}) + \bm{j}^{Q}(\bm{r}) \left[1 + \chi(\bm{r}) \right]^2}{2}
    -
    \frac{\left[1 + \chi(\bm{r}) \right]^2 \left[\nabla\times\bm{\Lambda}(\bm{r})\right]+\left[\nabla\times\bm{\Lambda}(\bm{r})\right]\left[1 + \chi(\bm{r}) \right]^2}{2}
    \nonumber\\
    &+
    \nabla\times\frac{\left[1 + \chi(\bm{r}) \right]^2 \bm{\Lambda}(\bm{r}) + \bm{\Lambda}(\bm{r})\left[1 + \chi(\bm{r}) \right]^2}{2},
    \label{eq:jQchi}
\end{align}
where $\bm{\Lambda} (\bm{r})$ is given by
\begin{align}
        \bm{\Lambda} (\bm{r}) = \frac{\hbar}{16i} \mathcal{A}^\dagger (\bm{r}) \left( \hat{\bm{v}} \times \sigma_3 \hat{\bm{v}} \right)
    \mathcal{A}^{}(\bm{r}).
\end{align}
By redefining
$\bm{j}^{Q;\chi} (\bm{r})
    -\nabla\times\left\{\left[1 + \chi(\bm{r}) \right]^2 \bm{\Lambda}(\bm{r}) + \bm{\Lambda}(\bm{r})\left[1 + \chi(\bm{r}) \right]^2\right\}/2$ and
$\bm{j}^{Q}(\bm{r}) - \nabla \times \bm{\Lambda}(\bm{r})$
as $\bm{j}^{Q;\chi} (\bm{r})$ and $\bm{j}^{Q}(\bm{r})$ respectively in Eq.~\eqref{eq:jQchi}, we find that the new thermal current operators satisfy the scaling relation given in Eq.~\eqref{appeq:scaling}.
Thus, the thermal current density is written as
\begin{align}
    \bm{j}^{Q}(\bm{r}) 
    = 
    \frac{1}{4} \mathcal{A}^\dagger (\bm{r}) 
    \left(
    \hat{\bm{v}} \sigma_3 \hat{\mathcal{M}}_{0} + \hat{\mathcal{M}}_{0} \sigma_3 \hat{\bm{v}}
    \right)
    \mathcal{A}^{}(\bm{r})
    - \frac{\hbar}{16i}\sum_{\lambda} \nabla_{\lambda}
    \left[ \mathcal{A}^\dagger (\bm{r}) \left( \hat{\bm{v}}\sigma_3 \hat{v}_{\lambda} - \hat{v}_{\lambda}\sigma_3 \hat{\bm{v}} \right) \mathcal{A}^{}(\bm{r}) \right].
\end{align}
Finally, we obtain Eq.~\eqref{eq:scaling-jq} by introducing the Fourier transformations of $\mathcal{M}_{\bm{\delta}}, \bm{j}^{Q} (\bm{r})$, and $\mathcal{A}^{}(\bm{r})$ as, 
\begin{align}
    \mathcal{M}_{\bm{q}} 
    = \sum_{\bm{\delta}} \mathcal{M}_{\bm{\delta}} e^{i\bm{q}\cdot \bm{\delta}},\quad
    \bm{j}^{Q}_{\bm{q}} = \int d\bm{r} \bm{j}^{Q} (\bm{r}) e^{-i\bm{q}\cdot \bm{r}},\quad
    \mathcal{A}^{}_{\bm{q}} = \int d\bm{r} \mathcal{A}^{}(\bm{r}) e^{-i\bm{q}\cdot \bm{r}}.
\end{align}

\section{Expression of transport coefficient}
\label{app:transport-coefficient}

\subsection{Expression of $S_{xy}$}
\label{app:express-Sxy}

In this section, we present the detailed derivation of $S_{\lambda\lambda'}$ given in Eq.~\eqref{eq:Sxy-green}.
The current-current correlation in Eq.~\eqref{eq:Piw} is written by the sum of four products of the bosonic operators $\mathcal{A}$ and $\mathcal{A}^\dagger$ by using the expression of $\bm{J}^{Q}$ in Eq.~\eqref{eq:JQ-form}.
We apply the following decouplings to them, which corresponds to neglecting vertex corrections:
\begin{align}
    P_{\lambda\lambda'} (i\Omega) \simeq
    -\frac{1}{16V} \int_{0}^{\beta} d\tau e^{i\Omega\tau} \sum_{s_1 s_2 s_3 s_4=1}^{2N}\sum_{\bm{k}\bm{k}'}
    \big(X_{\bm{k},\lambda}\big)_{s_1 s_2} \big(X_{\bm{k}',\lambda'}\big)_{s_3 s_4}
    \Bigg[&
        \left\langle T_{\tau} \mathcal{A}_{\bm{k},s_1}^\dagger (\tau) \mathcal{A}_{\bm{k}',s_4}^{} \right\rangle 
        \left\langle T_{\tau} \mathcal{A}_{\bm{k},s_2}^{}(\tau) \mathcal{A}_{\bm{k}',s_3}^\dagger \right\rangle 
        \nonumber\\
        &
        +
        \left\langle T_{\tau} \mathcal{A}_{\bm{k},s_1}^\dagger (\tau) \mathcal{A}_{\bm{k}',s_3}^{\dagger} \right\rangle 
        \left\langle T_{\tau} \mathcal{A}_{\bm{k},s_2}^{}(\tau) \mathcal{A}_{\bm{k}',s_4}^{} \right\rangle 
    \Bigg],
\end{align}  
where we introduce
$
X_{\bm{k},\lambda} = v_{\bm{k},x}\sigma_3 \mathcal{M}_{\bm{k}} + \mathcal{M}_{\bm{k}}\sigma_3 v_{\bm{k},\lambda}
$.
The bosonic operators $\mathcal{A}$ and $\mathcal{A}^\dagger$ are written by using the Bogoliubov bosons given in Eq.~\eqref{eq:Bogoliubov-Bop} as
\begin{align}
    \label{appeq:AtoB1}
    \mathcal{A}_{\bm{k},s}^{} &= \sum_{\eta=1}^{N} \left(T_{\bm{k}}^{}\right)_{s\eta} b_{\bm{k},\eta}^{} + \sum_{\eta=1}^{N} \left(T_{\bm{k}}^{}\right)_{s,\eta+N} b_{-\bm{k},\eta}^\dagger,\\
    \label{appeq:AtoB2}
    \mathcal{A}_{\bm{k},s}^{\dagger} &= \sum_{\eta=1}^{N} \left(T_{\bm{k}}^\dagger\right)_{\eta s} b_{\bm{k},\eta}^{\dagger} + \sum_{\eta=1}^{N} \left(T_{\bm{k}}^\dagger\right)_{\eta+N, s} b_{-\bm{k},\eta}^{}.
\end{align}
By neglecting the off-diagonal part of the temperature Green's function for $\eta$ in Eq.~\eqref{eq:definition-Green}, we obtain the following form:
\begin{align}
    P_{\lambda\lambda'} (i\Omega) 
    \simeq 
    &-\frac{1}{8V} \int_{0}^{\beta} \sum_{\eta,\eta'=1}^{2N}\sum_{\bm{k}}
    \left( T_{\bm{k}}^\dagger X_{\bm{k}}^{\lambda} T_{\bm{k}}^{} \right)_{\eta\eta'} \left( T_{\bm{k}}^\dagger X_{\bm{k}}^{\lambda'} T_{\bm{k}}^{} \right)_{\eta'\eta} 
    \mathcal{G}_{\bm{k},\eta}^{} (-\tau) \mathcal{G}_{\bm{k},\eta'}^{} (\tau)\nonumber\\
    = 
    &-\frac{k_{\rm B} T}{8V} \sum_{n}^{\infty} \sum_{\eta,\eta'=1}^{2N}\sum_{\bm{k}}
    \left( T_{\bm{k}}^\dagger X_{\bm{k}}^{x} T_{\bm{k}}^{} \right)_{\eta\eta'} \left( T_{\bm{k}}^\dagger X_{\bm{k}}^{y} T_{\bm{k}}^{} \right)_{\eta'\eta} 
    \mathcal{G}_{\bm{k},\eta}^{} (i\omega_{n}-i\Omega) \mathcal{G}_{\bm{k},\eta'}^{} (i\omega_{n})
\end{align}
The Matsubara sum can be taken by performing the integrals along the three contours shown in Fig.~\ref{fig:matsubara} on the complex plane~\cite{eliashberg1962}.
Carrying out the analytic continuation for the Matsubara frequency, we obtain the retarded correlation function between thermal currents as 
\begin{align}
    P_{\lambda\lambda'}^{R} (\Omega) \simeq &
    -\frac{1}{8V} \sum_{\eta,\eta'=1}^{2N}\sum_{\bm{k}}
    \left( T_{\bm{k}}^\dagger X_{\bm{k},\lambda}^{} T_{\bm{k}}^{} \right)_{\eta\eta'} \left( T_{\bm{k}}^\dagger X_{\bm{k},\lambda'}^{} T_{\bm{k}}^{} \right)_{\eta'\eta} \nonumber\\
    &\times\mathrm{P}\int_{-\infty}^{\infty} \frac{d\omega}{2\pi i} g(\beta\omega)
    \Big[
    G_{\bm{k},\eta}^{R}(\omega) G_{\bm{k},\eta'}^{R} (\omega+\hbar\Omega) - G_{\bm{k},\eta}^{A}(\omega) G_{\bm{k},\eta'}^{R} (\omega+\hbar\Omega)
    +
    G_{\bm{k},\eta}^{A}(\omega-\hbar\Omega) G_{\bm{k},\eta'}^{R} (\omega) - G_{\bm{k},\eta}^{A}(\omega-\hbar\Omega) G_{\bm{k},\eta'}^{A} (\omega)
    \Big]\nonumber\\
    = &
    -\frac{1}{8V} \sum_{\eta,\eta'=1}^{2N}\sum_{\bm{k}}
    \left( T_{\bm{k}}^\dagger v_{\bm{k},\lambda}^{} T_{\bm{k}}^{} \right)_{\eta\eta'} \left( T_{\bm{k}}^\dagger v_{\bm{k},\lambda'}^{} T_{\bm{k}}^{} \right)_{\eta'\eta} 
    \left( \sigma_{3,\eta}{\mathcal{E}}_{\bm{k},\eta} + \sigma_{3,\eta'}{\mathcal{E}}_{\bm{k},\eta'} \right)^2\nonumber\\
    &\times\mathrm{P}\int_{-\infty}^{\infty} \frac{d\omega}{2\pi i} g(\beta\omega)
    \Big[
    G_{\bm{k},\eta}^{R}(\omega) G_{\bm{k},\eta'}^{R} (\omega+\hbar\Omega) - G_{\bm{k},\eta}^{A}(\omega) G_{\bm{k},\eta'}^{R} (\omega+\hbar\Omega)
    +
    G_{\bm{k},\eta}^{A}(\omega-\hbar\Omega) G_{\bm{k},\eta'}^{R} (\omega) - G_{\bm{k},\eta}^{A}(\omega-\hbar\Omega) G_{\bm{k},\eta'}^{A} (\omega)
    \Big],
\end{align}
where we use the relation 
$\left(T_{\bm{k}}^\dagger X_{\bm{k},\lambda}^{} T_{\bm{k}}^{}  \right)_{\eta\eta'}
=
\left( \sigma_{3,\eta}{\mathcal{E}}_{\bm{k},\eta} + \sigma_{3,\eta'}{\mathcal{E}}_{\bm{k},\eta'} \right) \left(T_{\bm{k}}^\dagger v_{\bm{k},\lambda}^{} T_{\bm{k}}^{} \right)_{\eta\eta'}
$, which is calculated from 
$T_{\bm{k}}^{\dagger} M_{\bm{k}} \sigma_3 = \mathcal{E}_{\bm{k}}\sigma_3 T_{\bm{k}}^\dagger$.
Finally, substituting the above equation to Eq.~\eqref{eq:def-Sxy}, we obtain Eq.~\eqref{eq:Sxy-green}.

\begin{figure}[b]
  \begin{center}
    \includegraphics[width=70mm,clip]{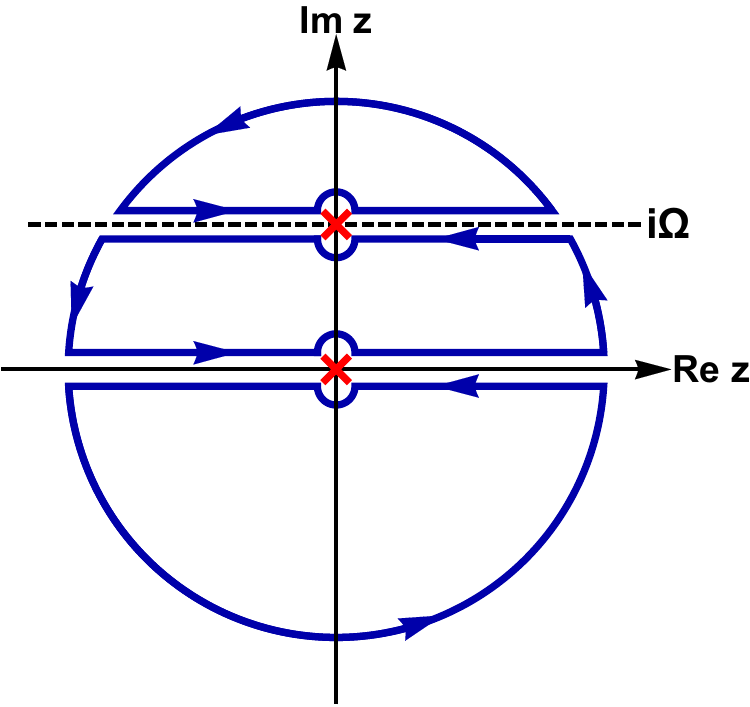}
    \caption{
    Paths of the contour integrals for $F(z)=  g(\beta z) \mathcal{G}_{\bm{k},\eta}(z-i\Omega) \mathcal{G}_{\bm{k},\eta'}(z)$.
    The horizontal dashed line represents $\mathrm{Im}z = \Omega$.
    }
    \label{fig:matsubara}
  \end{center}
\end{figure}

\subsection{Expression of  $\tilde{M}_{z}^{Q}$}
\label{app:express-tildeMz}

Next, we derive the expression of $\tilde{M}_{z}^{Q}$ in Eq.~\eqref{eq:tildeMz-green}.
By carrying out calculations similar to the procedure obtaining $P_{\lambda\lambda'}(i\Omega)$ for $\langle h_{-\bm{q}};j_{\bm{q},\lambda}\rangle$ with Eqs.~\eqref{eq:hq-def} and \eqref{eq:scaling-jq}, we obtain the following result:
\begin{align}
    \left\langle h_{-\bm{q}};j_{\bm{q},\lambda}\right\rangle
    \simeq
    &\frac{1}{4\beta} \sum_{\eta,\eta'=1}^{2N} \sum_{\bm{k}} 
    \left(T_{\bm{k}}^\dagger \frac{\mathcal{M}_{\bm{k}}+\mathcal{M}_{\bm{k}-\bm{q}}}{2} T_{\bm{k}-\bm{q}}^{}\right)_{\eta\eta'}
    \left(T_{\bm{k}-\bm{q}}^\dagger Y_{\bm{k}-\bm{q},\bm{k},\lambda}^{} T_{\bm{k}}^{} \right)_{\eta'\eta}
    \nonumber\\
    \label{appeq:hj-green}
    & \qquad\qquad
    \times\mathrm{P}\int_{-\infty}^{\infty} \frac{d\omega}{\pi}  g(\beta\omega)
    \Bigg\{
      \mathrm{Im} \left[G_{\bm{k},\eta}^{R} (\omega)\right] G_{\bm{k-\bm{q}},\eta'}^{R}(\omega)
      + G_{\bm{k},\eta}^{A}(\omega) \mathrm{Im}\left[G_{\bm{k}-\bm{q},\eta'}^{R}(\omega)\right]
    \Bigg\},
\end{align}
where $Y_{\bm{k}-\bm{q},\bm{k},\lambda}^{}$ is defined as
\begin{align}
    Y_{\bm{k}-\bm{q},\bm{k},\lambda}^{} = &
    \left( v_{\bm{k}-\bm{q},\lambda}^{} \sigma_3^{} \mathcal{M}_{\bm{k}}^{} + \mathcal{M}_{\bm{k}-\bm{q}}^{} \sigma_3^{} v_{\bm{k},\lambda}^{} \right)
    - \frac{1}{4} \sum_{\lambda'} 
    \hbar q_{\lambda'}^{}
    \left( v_{\bm{k}-\bm{q},\lambda}^{} \sigma_3^{} v_{\bm{k},\lambda'}^{} - v_{\bm{k}-\bm{q},\lambda'}^{} \sigma_3^{} v_{\bm{k},\lambda}^{} \right).
\end{align}
Substituting this expression to Eq.~\eqref{eq:def-tildeMz}, we obtain Eq.~\eqref{eq:tildeMz-green}.

\section{Evaluation of thermal Hall conductivity}
\label{app:kxy}
\subsection{Calculation of $S_{\lambda\lambda'}$}
\label{app:calc-Sxy}

In this section, we derive Eq.~\eqref{eq:Sxy}.
By substituting Eqs.~\eqref{eq:interaction GreenR-appro} and \eqref{eq:interaction GreenA-appro} to Eq.~\eqref{eq:Sxy-green}, $S_{\lambda\lambda'}$ is written as
\begin{align}
    S_{\lambda\lambda'}
     \simeq &-\frac{i\hbar}{8V} \sum_{\eta,\eta'=1}^{2N} \sum_{\bm{k}} \sigma_{3,\eta} \sigma_{3,\eta'} \left( \sigma_{3,\eta} \mathcal{E}_{\bm{k},\eta} + \sigma_{3,\eta'}{\mathcal{E}}_{\bm{k},\eta'} \right)^2
    \left( T_{\bm{k}}^\dagger v_{\bm{k},\lambda}^{} T_{\bm{k}}^{} \right)_{\eta\eta'} \left( T_{\bm{k}}^\dagger v_{\bm{k},\lambda'}^{} T_{\bm{k}}^{} \right)_{\eta'\eta} \nonumber\\
    &\qquad\qquad\times\mathrm{P}\int_{-\infty}^{\infty} d\omega
    \left[
        \sigma_{3,\eta}\rho_{\bm{k},\eta} (\omega) \frac{g(\beta\omega)}{\left(\omega - \sigma_{3,\eta'}\mathcal{E}_{\bm{k},\eta'}+i\sigma_{3,\eta'}\Gamma_{\bm{k},\eta'}(\omega)\right)^2}
        -
        \sigma_{3,\eta'}\rho_{\bm{k},\eta'} (\omega) \frac{g(\beta\omega)}{\left(\omega - \sigma_{3,\eta}\mathcal{E}_{\bm{k},\eta}-i\sigma_{3,\eta}\Gamma_{\bm{k},\eta}(\omega)\right)^2}
    \right]\nonumber\\
    =&\frac{\hbar}{4V} \sum_{\eta,\eta'=1}^{2N} \sum_{\bm{k}} \sigma_{3,\eta} \sigma_{3,\eta'} \left( \sigma_{3,\eta}\mathcal{E}_{\bm{k},\eta} + \sigma_{3,\eta'}\mathcal{E}_{\bm{k},\eta'} \right)^2
    \mathrm{Im}
    \left[
    \left( T_{\bm{k}}^\dagger v_{\bm{k},\lambda}^{} T_{\bm{k}}^{} \right)_{\eta\eta'} \left( T_{\bm{k}}^\dagger v_{\bm{k},\lambda'}^{} T_{\bm{k}}^{} \right)_{\eta'\eta}
    \mathrm{P}\int_{-\infty}^{\infty} d\omega
    \frac{\sigma_{3,\eta}\rho_{\bm{k},\eta} (\omega) g(\beta\omega)}{\left(\omega - \sigma_{3,\eta'}\mathcal{E}_{\bm{k},\eta'}+i\sigma_{3,\eta'}\Gamma_{\bm{k},\eta'}(\omega)\right)^2}
    \right].
    \end{align}
Here, we introduce the symmetric and antisymmetic parts of the above expression with respect to $(\lambda,\lambda')$ as $S_{\lambda\lambda'}^{s}$ and $S_{\lambda\lambda'}^{a}$, respectively, which are represented as
    \begin{align}
    S_{\lambda\lambda'}^{s} = &\frac{\hbar}{4V} \sum_{\eta,\eta'=1}^{2N}\sum_{\bm{k}}
    \sigma_{3,\eta}\sigma_{3,\eta'}
    \left( \sigma_{3,\eta}\mathcal{E}_{\bm{k},\eta} + \sigma_{3,\eta'}\mathcal{E}_{\bm{k},\eta'} \right)^2
    \nonumber\\
    &\qquad\qquad\times
    \mathrm{Re}
    \left[     
    \left( T_{\bm{k}}^\dagger v_{\bm{k},\lambda}^{} T_{\bm{k}}^{} \right)_{\eta\eta'} \left( T_{\bm{k}}^\dagger v_{\bm{k},\lambda'}^{} T_{\bm{k}}^{} \right)_{\eta'\eta}
    \right]
    \mathrm{Im}
    \left[
       \mathrm{P}\int_{-\infty}^{\infty} d\omega
        \frac{\sigma_{3,\eta}\rho_{\bm{k},\eta} (\omega)g(\beta\omega)}{\left(\omega - \sigma_{3,\eta'}\mathcal{E}_{\bm{k},\eta'} + i\sigma_{3,\eta'}\Gamma_{\bm{k},\eta'}(\omega)\right)^2}
    \right],\\
    S_{\lambda\lambda'}^{a} = &\frac{\hbar}{4V} \sum_{\eta,\eta'=1}^{2N}\sum_{\bm{k}}
    \sigma_{3,\eta}\sigma_{3,\eta'}
    \left( \sigma_{3,\eta}\mathcal{E}_{\bm{k},\eta} + \sigma_{3,\eta'}\mathcal{E}_{\bm{k},\eta'} \right)^2 
    \nonumber\\
    &\qquad\qquad\times
    \mathrm{Im}
    \left[     
    \left( T_{\bm{k}}^\dagger v_{\bm{k},\lambda}^{} T_{\bm{k}}^{} \right)_{\eta\eta'} \left( T_{\bm{k}}^\dagger v_{\bm{k},\lambda'}^{} T_{\bm{k}}^{} \right)_{\eta'\eta}
    \right]
    \mathrm{Re}
    \left[
       \mathrm{P}\int_{-\infty}^{\infty} d\omega
        \frac{\sigma_{3,\eta}\rho_{\bm{k},\eta} (\omega)g(\beta\omega)}{\left(\omega - \sigma_{3,\eta'}\mathcal{E}_{\bm{k},\eta'} + i\sigma_{3,\eta'}\Gamma_{\bm{k},\eta'}(\omega)\right)^2}
    \right].\label{appeq:Santi}
\end{align}
As we calculate thermal Hall conductivity, we focus on the antisymmetric part $S_{\lambda\lambda'}^{a}$.
Using $\tilde{\Omega}_{\bm{k},\eta\eta'}^{\lambda}$ introduced in Eq.~\eqref{eq:Omega-tilde}, we rewrite Eq.~\eqref{appeq:Santi} as 
\begin{align}
    S_{\lambda\lambda'}^{a}
    = &\frac{1}{4\hbar V} \sum_{\lambda''}\varepsilon_{\lambda\lambda'\lambda''}
    \sum_{\eta,\eta'=1}^{2N}\sum_{\bm{k}} \tilde{\Omega}_{\bm{k},\eta\eta'}^{\lambda''}
    \left( \sigma_{3,\eta}\mathcal{E}_{\bm{k},\eta} + \sigma_{3,\eta'}\mathcal{E}_{\bm{k},\eta'} \right)^2 \left( \sigma_{3,\eta}\mathcal{E}_{\bm{k},\eta} - \sigma_{3,\eta'}\mathcal{E}_{\bm{k},\eta'} \right)^2
    \nonumber\\
    &\qquad\qquad\qquad\qquad\times
    \mathrm{Re}
    \left[
       \mathrm{P}\int_{-\infty}^{\infty} d\omega
        \frac{\sigma_{3,\eta}\rho_{\bm{k},\eta} (\omega)g(\beta\omega)}{\left(\omega - \sigma_{3,\eta'}\mathcal{E}_{\bm{k},\eta'}+i\sigma_{3,\eta'}\Gamma_{\bm{k},\eta'}(\omega)\right)^2}
    \right]\nonumber\\
    \label{appeq:Sxy}
    = &\frac{1}{4\hbar V} \sum_{\lambda''}\varepsilon_{\lambda\lambda'\lambda''}
    \sum_{\eta=1}^{N}\sum_{\eta'=1}^{2N}\sum_{\bm{k}} \tilde{\Omega}_{\bm{k},\eta\eta'}^{\lambda''}
    \left( \varepsilon_{\bm{k},\eta} + \sigma_{3,\eta'}\mathcal{E}_{\bm{k},\eta'} \right)^2 \left( \varepsilon_{\bm{k},\eta} - \sigma_{3,\eta'}\mathcal{E}_{\bm{k},\eta'} \right)^2
    \nonumber\\
    &\qquad\qquad\qquad\qquad\times
    \mathrm{Re}
    \left[
       \mathrm{P}\int_{-\infty}^{\infty} d\omega
        \rho_{\bm{k},\eta} (\omega) \frac{2g(\beta\omega) + 1}{\left(\omega - \sigma_{3,\eta'}\mathcal{E}_{\bm{k},\eta'}+i\sigma_{3,\eta'}\Gamma_{\bm{k},\eta'}(\omega)\right)^2}
    \right],
\end{align}
which is the same as Eq.~\eqref{eq:Sxy}.
Here,
we use
    $\rho_{-\bm{k},\eta+N} (-\omega) = -\rho_{\bm{k},\eta} (\omega)$
    and $
    \tilde{\Omega}_{-\bm{k},\eta+N,\eta'+N}^{\lambda} = -\tilde{\Omega}_{\bm{k},\eta\eta'}^{\lambda}$ for $\eta,\eta' = 1,\cdots,N$ and the relation for $\eta\neq\eta'$ as follows:
\begin{align}
    \label{eq:TVT}
    \left(T_{\bm{k}}^\dagger v_{\bm{k},\lambda}^{} T_{\bm{k}}^{} \right)_{\eta\eta'}
    =
    \frac{1}{\hbar} 
    \left( \sigma_{3,\eta'}\mathcal{E}_{\bm{k},\eta'}^{} - \sigma_{3,\eta}\mathcal{E}_{\bm{k},\eta}^{} \right) \left( T_{\bm{k}}^\dagger \sigma_3 \pdiff{T_{\bm{k}}^{}}{k_{\lambda}} \right)_{\eta\eta'}.
\end{align}

\subsection{Calculation of $M_{\lambda}^{Q}$}
\label{app:calc-Mz}

In this section, we derive the expression of $M_{\lambda}^{Q}$ in Eq.~\eqref{eq:Mz}.
Similar to the previous section, we substitute the Green's functions given in Eqs.~\eqref{eq:interaction GreenR-appro} and \eqref{eq:interaction GreenA-appro} to Eq.~\eqref{eq:tildeMz-green}, and thereby, we obtain the following form:
\begin{align}
    \tilde{M}_{\lambda}^{Q} \simeq & \frac{1}{16i} \sum_{\lambda'\lambda''} \varepsilon_{\lambda\lambda'\lambda''}\pdiff{}{q_{\lambda''}}
    \sum_{\eta,\eta'=1}^{2N} \sum_{\bm{k}} \sigma_{3,\eta}\sigma_{3,\eta'}
    \left[ T_{\bm{k}}^\dagger \left(\mathcal{M}_{\bm{k}}^{}+\mathcal{M}_{\bm{k}-\bm{q}}^{}\right) T_{\bm{k}-\bm{q}}^{} \right]_{\eta\eta'}
    \left( T_{\bm{k}-\bm{q}}^{} Y_{\bm{k}-\bm{q},\bm{k},x}^{} T_{\bm{k}}^{} \right)_{\eta'\eta}\nonumber\\
    &\qquad\qquad\qquad\qquad
    \times\left.
    \mathrm{P}\int_{-\infty}^{\infty} d\omega
    \left[
        \frac{\sigma_{3,\eta}\rho_{\bm{k},\eta}(\omega)g(\beta\omega)}{\omega-\sigma_{3,\eta'}\mathcal{E}_{\bm{k}-\bm{q},\eta'} + i\sigma_{3,\eta'}\Gamma_{\bm{k}-\bm{q},\eta'}(\omega)}
        +
        \frac{\sigma_{3,\eta'}\rho_{\bm{k}-\bm{q},\eta'}(\omega)g(\beta\omega)}{\omega-\sigma_{3,\eta}\mathcal{E}_{\bm{k},\eta} - i\sigma_{3,\eta}\Gamma_{\bm{k},\eta}(\omega)}
    \right]
    \right|_{\bm{q}\to \bm{0}}.
\end{align}
Here, we divide the above expression into two parts for $\eta \neq \eta'$ and $\eta = \eta'$, which are defined as $\tilde{M}_{\lambda}^{Q;\mathrm{inter}}$ and $\tilde{M}_{\lambda}^{Q;\mathrm{intra}}$, respectively.
First, we focus on $\tilde{M}_{\lambda}^{Q;\mathrm{inter}}$.
This is calculated as 
\begin{align}
    \tilde{M}_{\lambda}^{Q;\mathrm{inter}}&=
    \frac{1}{16i} \sum_{\lambda'\lambda''} \varepsilon_{\lambda\lambda'\lambda''}\pdiff{}{q_{\lambda''}}
    \sum_{\eta\ne \eta'}^{2N} \sum_{\bm{k}} \sigma_{3,\eta}\sigma_{3,\eta'}
    \left[ T_{\bm{k}}^\dagger \left(\mathcal{M}_{\bm{k}}^{}+\mathcal{M}_{\bm{k}-\bm{q}}^{}\right) T_{\bm{k}-\bm{q}}^{} \right]_{\eta\eta'}
    \left( T_{\bm{k}-\bm{q}}^{} Y_{\bm{k}-\bm{q},\bm{k},x}^{} T_{\bm{k}}^{} \right)_{\eta'\eta}\nonumber\\
    &\qquad\qquad\qquad\qquad
    \times \left.
    \mathrm{P}\int_{-\infty}^{\infty} d\omega
    \left[
        \frac{\sigma_{3,\eta}\rho_{\bm{k},\eta}(\omega)g(\beta\omega)}{\omega-\sigma_{3,\eta'}\mathcal{E}_{\bm{k}-\bm{q},\eta'} + i\sigma_{3,\eta'}\Gamma_{\bm{k}-\bm{q},\eta'}(\omega)}
        +
        \frac{\sigma_{3,\eta'}\rho_{\bm{k}-\bm{q},\eta'}(\omega)g(\beta\omega)}{\omega-\sigma_{3,\eta}\mathcal{E}_{\bm{k},\eta} - i\sigma_{3,\eta}\Gamma_{\bm{k},\eta}(\omega)}
    \right]
    \right|_{\bm{q}\to \bm{0}}
    \notag\\
    &\simeq-\frac{1}{16i}
    \sum_{\lambda'\lambda''}\varepsilon_{\lambda\lambda'\lambda''}\sum_{\eta\neq\eta'}^{2N}\sum_{\bm{k}} \sigma_{3,\eta} \sigma_{3,\eta'}
    \left( \sigma_{3,\eta}\mathcal{E}_{\bm{k},\eta} + \sigma_{3,\eta'}\mathcal{E}_{\bm{k},\eta'} \right)^2 \left( T_{\bm{k}}^\dagger \sigma_3 \pdiff{T_{\bm{k}}^{}}{k_{\lambda''}} \right)_{\eta\eta'}
    \left( T_{\bm{k}}^\dagger v_{\bm{k},\lambda'}^{} T_{\bm{k}}^{} \right)_{\eta'\eta}
    \nonumber\\
    &\qquad\qquad\qquad\qquad
    \times 
    \mathrm{P}\int_{-\infty}^{\infty} d\omega
    \left[
        \frac{\sigma_{3,\eta}\rho_{\bm{k},\eta}(\omega)g(\beta\omega)}{\omega-\sigma_{3,\eta'}\mathcal{E}_{\bm{k},\eta'} + i\sigma_{3,\eta'}\Gamma_{\bm{k},\eta'}(\omega)}
        +
        \frac{\sigma_{3,\eta'}\rho_{\bm{k},\eta'}(\omega)g(\beta\omega)}{\omega-\sigma_{3,\eta}\mathcal{E}_{\bm{k},\eta} - i\sigma_{3,\eta}\Gamma_{\bm{k},\eta}(\omega)}
    \right]\notag\\
        \label{eq:tildeMz-inter}
    &=
    -\frac{1}{4\hbar} \sum_{\eta,\eta'=1}^{2N}\sum_{\bm{k}} \tilde{\Omega}_{\bm{k},\eta\eta'}^{\lambda}
    \left( \sigma_{3,\eta}\mathcal{E}_{\bm{k},\eta} + \sigma_{3,\eta'}\mathcal{E}_{\bm{k},\eta'} \right)^2
    \left(\sigma_{3,\eta}\mathcal{E}_{\bm{k},\eta} - \sigma_{3,\eta'}\mathcal{E}_{\bm{k},\eta'}\right)
    \mathrm{Re}
    \left[
    \mathrm{P}\int_{-\infty}^{\infty} d\omega
        \frac{\sigma_{3,\eta}\rho_{\bm{k},\eta}(\omega)g(\beta\omega)}{\omega-\sigma_{3,\eta'}\mathcal{E}_{\bm{k},\eta'} + i\sigma_{3,\eta'}\Gamma_{\bm{k},\eta'}(\omega)}
    \right],
\end{align}
where we neglect the first derivative of $\Gamma_{\bm{k},\eta}(\omega)$ with respect to $\bm{k}$ and use the relations $\left( T_{\bm{k}}^\dagger \mathcal{M}_{\bm{k}}^{} T_{\bm{k}}^{}\right)_{\eta\eta'}=0$ and $\left( T_{\bm{k}}^\dagger v_{\bm{k},\lambda}^{} T_{\bm{k}}^{}\right)_{\eta\eta'}=\frac{1}{\hbar}\left(\sigma_{3,\eta'}\mathcal{E}_{\bm{k},\eta'} - \sigma_{3,\eta}\mathcal{E}_{\bm{k},\eta}\right)\left( T_{\bm{k}}^\dagger \sigma_3 \pdiff{T_{\bm{k}}^{}}{k_{\lambda}} \right)_{\eta\eta'}$ for $\eta \neq \eta'$.

Next, we calculate $\tilde{M}_{\lambda}^{Q;\mathrm{intra}}$ for the case with $\eta=\eta'$ as follows:
\begin{align}
    \tilde{M}_{\lambda}^{Q;\mathrm{intra}} 
    = & \frac{1}{16i}  \sum_{\lambda'\lambda''}
    \varepsilon_{\lambda\lambda'\lambda''} \pdiff{}{q_{\lambda''}}
    \sum_{\eta=1}^{2N} \sum_{\bm{k}}
    \left[ T_{\bm{k}}^\dagger \left(\mathcal{M}_{\bm{k}}^{}+\mathcal{M}_{\bm{k}-\bm{q}}^{}\right) T_{\bm{k}-\bm{q}}^{} \right]_{\eta\eta}
    \left( T_{\bm{k}-\bm{q}}^{} Y_{\bm{k}-\bm{q},\bm{k},\lambda'}^{} T_{\bm{k}}^{} \right)_{\eta\eta}
    \nonumber\\
    &\qquad\qquad\qquad\qquad
    \times 
    \left.
    \mathrm{P}\int_{-\infty}^{\infty} d\omega
    \left[
        \frac{\sigma_{3,\eta}\rho_{\bm{k},\eta}(\omega)g(\beta\omega)}{\omega-\sigma_{3,\eta}\mathcal{E}_{\bm{k}-\bm{q},\eta} + i\sigma_{3,\eta}\Gamma_{\bm{k}-\bm{q},\eta}(\omega)}
        +
        \frac{\sigma_{3,\eta}\rho_{\bm{k}-\bm{q},\eta}(\omega)g(\beta\omega)}{\omega-\sigma_{3,\eta}\mathcal{E}_{\bm{k},\eta} - i\sigma_{3,\eta}\Gamma_{\bm{k},\eta}(\omega)}
    \right]
    \right|_{\bm{q}\to \bm{0}}\nonumber\\
    = & -\frac{1}{4}
    \sum_{\lambda'\lambda''}
    \varepsilon_{\lambda\lambda'\lambda''}
    \sum_{\eta=1}^{2N} \sum_{\bm{k}}
    \mathcal{E}_{\bm{k},\eta}^2 
    \mathrm{Im}
    \left[ \left( T_{\bm{k}}^\dagger \sigma_3 \pdiff{T_{\bm{k}}^{}}{k_{\lambda''}}\right)_{\eta\eta} \left( T_{\bm{k}}^\dagger v_{\bm{k},\lambda'}^{} T_{\bm{k}}^{} \right)_{\eta\eta} \right]
    \mathrm{Re}
    \left[
    \mathrm{P}\int_{-\infty}^{\infty} d\omega
        \frac{2\sigma_{3,\eta}\rho_{\bm{k},\eta}(\omega)g(\beta\omega)}{\omega-\sigma_{3,\eta}\mathcal{E}_{\bm{k},\eta} + i\sigma_{3,\eta}\Gamma_{\bm{k},\eta}(\omega)}
    \right]\nonumber\\
    & -\frac{1}{8}
    \sum_{\lambda'\lambda''}
    \varepsilon_{\lambda\lambda'\lambda''}
    \sum_{\eta=1}^{2N} \sum_{\bm{k}}
    \mathcal{E}_{\bm{k},\eta}
    \mathrm{Im}
    \left\{
    \left[ \pdiff{T_{\bm{k}}^\dagger}{k_{\lambda''}} \left( \mathcal{M}_{\bm{k}}^{} \sigma_3 v_{\bm{k},\lambda'}^{} + v_{\bm{k},\lambda'}^{} \sigma_3 \mathcal{M}_{\bm{k}}^{} \right) T_{\bm{k}}^{} \right]_{\eta\eta}
    \right\}
    \mathrm{Re}
    \left[
    \mathrm{P}\int_{-\infty}^{\infty} d\omega
        \frac{2\sigma_{3,\eta}\rho_{\bm{k},\eta}(\omega)g(\beta\omega)}{\omega-\sigma_{3,\eta}\mathcal{E}_{\bm{k},\eta} + i\sigma_{3,\eta}\Gamma_{\bm{k},\eta}(\omega)}
    \right]\nonumber\\
    & -\frac{\hbar}{16}
    \sum_{\lambda'\lambda''}
    \varepsilon_{\lambda\lambda'\lambda''}
    \sum_{\eta=1}^{2N} \sum_{\bm{k}}
    \mathcal{E}_{\bm{k},\eta}
    \mathrm{Im}
    \left[ 
      \left( T_{\bm{k}}^{} v_{\bm{k},\lambda''}^{} \sigma_3 v_{\bm{k},\lambda'}^{} T_{\bm{k}}^{} \right)_{\eta\eta}
    \right]
    \mathrm{Re}
    \left[
    \mathrm{P}\int_{-\infty}^{\infty} d\omega
        \frac{2\sigma_{3,\eta}\rho_{\bm{k},\eta}(\omega)g(\beta\omega)}{\omega-\sigma_{3,\eta}\mathcal{E}_{\bm{k},\eta} + i\sigma_{3,\eta}\Gamma_{\bm{k},\eta}(\omega)}
    \right].
\end{align}
Using the relation $\sigma_3 = T_{\bm{k}}^\dagger \sigma_3 T_{\bm{k}}^{} = T_{\bm{k}}^{} \sigma_3 T_{\bm{k}}^{\dagger}$ and Eq.~\eqref{eq:TVT}, we rewrite the above form as
\begin{align}
    \tilde{M}_{\lambda}^{Q;\mathrm{intra}}
    =&
    -\frac{1}{16\hbar}\sum_{\lambda'\lambda''} \varepsilon_{\lambda\lambda'\lambda''} \sum_{\eta,\eta'=1}^{2N} \sigma_{3,\eta}\mathcal{E}_{\bm{k},\eta}^{}
    \mathrm{Im}
    \left\{
      \left( \sigma_3 \pdiff{T_{\bm{k}}^\dagger}{k_{\lambda'}} \sigma_3 T_{\bm{k}}^{} \right)_{\eta\eta'}
      \left[ \left(\sigma_{3,\eta}{\mathcal{E}}_{\bm{k},\eta}^{} + \sigma_{3,\eta'}{\mathcal{E}}_{\bm{k},\eta'}^{} \right)^2 - 4 \sigma_{3,\eta}\mathcal{E}_{\bm{k},\eta}^2 \right]
      \left( \sigma_3 T_{\bm{k}}^\dagger \sigma_3 \pdiff{T_{\bm{k}}^{}}{k_{\lambda''}} \right)_{\eta'\eta}
    \right\}
        \nonumber\\
    &\qquad\qquad\qquad\qquad
    \times 
    \mathrm{Re}
    \left[ 
    \mathrm{P}\int_{-\infty}^{\infty} d\omega
        \frac{2\sigma_{3,\eta}\rho_{\bm{k},\eta}(\omega)g(\beta\omega)}{\omega-\sigma_{3,\eta}\mathcal{E}_{\bm{k},\eta} + i\sigma_{3,\eta}\Gamma_{\bm{k},\eta}(\omega)}
    \right].
\end{align}
By neglecting the $\omega$-derivative of $\Gamma_{\bm{k},\eta}(\omega)$, the $\omega$-integral in the above equation is approximated as 
\begin{align}
    \mathrm{Re}
    \left[
    \mathrm{P}\int_{-\infty}^{\infty} d\omega
        \frac{2\sigma_{3,\eta}\rho_{\bm{k},\eta}(\omega)g(\beta\omega)}{\omega-\sigma_{3,\eta}{\mathcal{E}}_{\bm{k},\eta} + i\sigma_{3,\eta}\Gamma_{\bm{k},\eta}(\omega)}
    \right]
    \simeq
    -\mathrm{P}\int_{-\infty}^{\infty} d\omega \sigma_{3,\eta}\pdiff{\rho_{\bm{k},\eta}}{\omega} g(\beta\omega)
    =
    \mathrm{P}\int_{-\infty}^{\infty}
    d\omega \sigma_{3,\eta}\rho_{\bm{k},\eta}(\omega) \pdiff{g}{\omega}.
\end{align}
Finally, $\tilde{M}_{\lambda}^{Q;\mathrm{intra}}$ is represented as
\begin{align}
    \label{eq:intra-tildeMz}
    \tilde{M}_{\lambda}^{Q;\mathrm{intra}} 
    \simeq
    -\frac{1}{8\hbar} \sum_{\eta,\eta'=1}^{2N} \sum_{\bm{k}} \tilde{\Omega}_{\bm{k},\eta\eta'}^{\lambda} \sigma_{3,\eta}\mathcal{E}_{\bm{k},\eta}
    \left[ \left(\sigma_{3,\eta}\mathcal{E}_{\bm{k},\eta}^{} + \sigma_{3,\eta'}\mathcal{E}_{\bm{k},\eta'}^{} \right)^2 - 4 \left(\sigma_{3,\eta}\mathcal{E}_{\bm{k},\eta}\right)^2 \right]
    \mathrm{P}\int_{-\infty}^{\infty}
    d\omega \sigma_{3,\eta}\rho_{\bm{k},\eta}(\omega) \pdiff{g}{\omega},
\end{align}
where the temperature derivative of $\Gamma_{\bm{k},\eta}(\omega)$ is neglected.
From Eqs.\eqref{eq:tildeMz-inter} and \eqref{eq:intra-tildeMz}, we obtain $\tilde{M}_{\lambda}^{Q}$, and $M_{\lambda}^{Q}$ in Eq.~\eqref{eq:Mz} is derived by solving the differential equation in Eq.~\eqref{eq:def-Mz} under the boundary condition $\lim_{\beta\to\infty}\beta \pdiff{\bm{M}^{Q}}{\beta} = 0$.

\subsection{Calculation of $\kappa_{\lambda\lambda'}^{H}$}
\label{app:calc-kxy}

In this section, we derive the expression of the thermal Hall conductivity in Eq.~\eqref{eq:kxy-A}.
First, we divide Eq~\eqref{eq:Mz} into to parts as $M_{\lambda}^{Q} \simeq M_{\lambda}^{Q;\mathrm{inter}} + M_{\lambda}^{Q;\mathrm{intra}}$, which are defined by
\begin{align}
    \frac{2M_{\lambda}^{Q;\mathrm{inter}}}{V}
    = &-\frac{1}{2\beta^{2}\hbar V}
    \sum_{\eta=1}^{N}\sum_{\eta'=1}^{2N} \sum_{\bm{k}}\tilde{\Omega}_{\bm{k},\eta\eta'}^{\lambda}
    \left( \varepsilon_{\bm{k},\eta} + \sigma_{3,\eta'}\mathcal{E}_{\bm{k},\eta'}^{} \right)^2
    \left( \varepsilon_{\bm{k},\eta} - \sigma_{3,\eta'}\mathcal{E}_{\bm{k},\eta'}^{} \right)
           \nonumber\\
    &\qquad\qquad\qquad\qquad
    \times 
    \mathrm{Re}
    \left[ \mathrm{P}
      \int_{-\infty}^{\infty} d\omega \frac{\rho_{\bm{k},\eta}(\omega)}{\omega - \sigma_{3,\eta'}\mathcal{E}_{\bm{k},\eta'} + i\sigma_{3,\eta'}\Gamma_{\bm{k},\eta'}(\omega)}
    \right]
    \int_{0}^{\beta} \tilde{\beta} \left[ 2g(\tilde{\beta}\omega) + 1 \right] d\tilde{\beta}\nonumber\\
    \label{appeq:M-inter}
    &-\frac{1}{2\beta^{2}\hbar V} 
    \sum_{\eta=1}^{N}\sum_{\eta'=1}^{2N} \sum_{\bm{k}} \tilde{\Omega}_{\bm{k},\eta\eta'}^{\lambda}
    \left( \varepsilon_{\bm{k},\eta} + \sigma_{3,\eta'}\mathcal{E}_{\bm{k},\eta'}^{} \right)^2  
    \mathrm{P}\int_{-\infty}^{\infty} d\omega \rho_{\bm{k},\eta} (\omega) \frac{\varepsilon_{\bm{k},\eta}}{\omega} 
    \int_{0}^{\beta} \tilde{\beta}\omega \pdiff{g}{\omega} d\tilde{\beta},
    \end{align}
    and
    \begin{align}
    \label{appeq:M-intra}
    \frac{2M_{\lambda}^{Q;\mathrm{intra}}}{V}
    = &
    -\frac{1}{\beta^{2}\hbar V} 
    \sum_{\eta=1}^{N} \sum_{\bm{k}}  \Omega_{\bm{k},\eta}^{\lambda} \varepsilon_{\bm{k},\eta}^3
    \mathrm{P}\int_{-\infty}^{\infty} d\omega \rho_{\bm{k},\eta} (\omega)
    \int_{0}^{\beta} \tilde{\beta} \pdiff{g}{\omega} d\tilde{\beta},
\end{align}
where we use the following relation for the Berry curvature given in Eq.~\eqref{eq:def-berry}~\cite{koyama2021}:
\begin{align}
    \label{appeq:berry-trans}
    \Omega_{\bm{k},\eta}^{\lambda}
    = -\sum_{\lambda'\lambda''}
    \varepsilon_{\lambda\lambda'\lambda''} \mathrm{Im} \left[ \sigma_3 \pdiff{T_{\bm{k}}^\dagger}{k_{\lambda'}} \sigma_3 \pdiff{T_{\bm{k}}^{}}{k_{\lambda''}} \right]_{\eta\eta}
    = 
    -\hbar^2\sum_{\lambda'\lambda''}\varepsilon_{\lambda\lambda'\lambda''} \sum_{\eta'(\neq\eta)}^{2N} 
    \frac{
    \sigma_{3,\eta}\sigma_{3,\eta'} 
    \mathrm{Im}\left[\left( T_{\bm{k}}^\dagger v_{\bm{k},\lambda'}^{} T_{\bm{k}}^{} \right)_{\eta\eta'} \left( T_{\bm{k}}^\dagger v_{\bm{k},\lambda''}^{} T_{\bm{k}}^{} \right)_{\eta'\eta}\right]
    }
    {\left( \sigma_{3,\eta}{\mathcal{E}}_{\bm{k},\eta} - \sigma_{3,\eta'}{\mathcal{E}}_{\bm{k},\eta'} \right)^2}
    = -2 \sum_{\eta'(\neq\eta)}^{2N}
    \tilde{\Omega}_{\bm{k},\eta\eta'}^{\lambda}.
\end{align}
We carry out the temperature integrals in Eqs.~\eqref{appeq:M-inter} and \eqref{appeq:M-intra} using the following relations:
\begin{align}
  \int_{0}^{\beta} \tilde{\beta} g(\tilde{\beta}\omega) d\tilde{\beta}
  &=
  \frac{1}{2} \beta^{2} g(\beta\omega) - \frac{1}{2\omega^2} \tilde{c}_2(g(\beta\omega)),\quad
  \int_{0}^{\beta} \tilde{\beta}\omega \pdiff{g}{\omega} d\tilde{\beta}
  = 
  \frac{1}{\omega^2} \tilde{c}_2(g(\beta\omega)).
\end{align}
Then, the thermal transport coefficients, $L_{\lambda\lambda'}^{a;\mathrm{inter}} = S_{\lambda\lambda'}^a + \sum_{\lambda''}
\varepsilon_{\lambda\lambda'\lambda''} \frac{2M_{\lambda''}^{Q;\mathrm{inter}}}{V}$ and
$L_{\lambda\lambda'}^{a;\mathrm{intra}} = 
\sum_{\lambda''}
\varepsilon_{\lambda\lambda'\lambda''}\frac{2M_{\lambda''}^{Q;\mathrm{intra}}}{V}$, where $L_{\lambda\lambda'}^{a}$ is written as $L_{\lambda\lambda'}^{a} \simeq L_{\lambda\lambda'}^{a;\mathrm{inter}}+L_{\lambda\lambda'}^{a;\mathrm{intra}}$, are calculated as
\begin{align}
    L_{\lambda\lambda'}^{a;\mathrm{inter}}
    =&
    -\frac{1}{2 \beta^{2}\hbar V} \sum_{\lambda''}\varepsilon_{\lambda\lambda'\lambda''}
    \sum_{\eta=1}^{N}\sum_{\eta'=1}^{2N}\sum_{\bm{k}} \tilde{\Omega}_{\bm{k},\eta\eta'}^{\lambda''}
    \left( \varepsilon_{\bm{k},\eta} + \sigma_{3,\eta'}\mathcal{E}_{\bm{k},\eta'} \right)^2 
    \left( \varepsilon_{\bm{k},\eta} - \sigma_{3,\eta'}\mathcal{E}_{\bm{k},\eta'}\right)\nonumber\\
    &
    \times\mathrm{Re}
    \left\{\mathrm{P}
    \int_{-\infty}^{\infty} d\omega \frac{\rho_{\bm{k},\eta}(\omega)}{\omega - \sigma_{3,\eta'}\mathcal{E}_{\bm{k},\eta'} + i \sigma_{3,\eta'}{\Gamma}_{\bm{k},\eta'}(\omega)}
    \left[
      1 - \frac{\varepsilon_{\bm{k},\eta} - \sigma_{3,\eta'}\mathcal{E}_{\bm{k},\eta'}}{\omega - \sigma_{3,\eta'}\mathcal{E}_{\bm{k},\eta'} + i \sigma_{3,\eta'}{\Gamma}_{\bm{k},\eta'}(\omega)}
    \right]
    \right\}
    \left\{\beta^{2}g(\beta\omega) + \frac{\beta^{2}}{2} - \frac{1}{\omega^2} \tilde{c}_2(g(\beta\omega))\right\}\nonumber\\
    &
    -\frac{1}{2\beta^{2}\hbar V} \sum_{\lambda''} \varepsilon_{\lambda\lambda'\lambda''}
    \sum_{\eta=1}^{N}\sum_{\eta'=1}^{2N} \sum_{\bm{k}} \tilde{\Omega}_{\bm{k},\eta\eta'}^{\lambda''} 
    \left(\varepsilon_{\bm{k},\eta} + \sigma_{3,\eta'}\mathcal{E}_{\bm{k},\eta'}\right)^2\nonumber\\
    \label{appeq:Lxy-inter}
    &\qquad\qquad
    \times\mathrm{Re}
    \left\{ \mathrm{P}
    \int_{-\infty}^{\infty} d\omega
    \frac{\rho_{\bm{k},\eta} (\omega)}{\omega^2}
    \left[
    \frac{\varepsilon_{\bm{k},\eta}}{\omega}
    - 
    \frac{\left(\varepsilon_{\bm{k},\eta} - \sigma_{3,\eta'}\mathcal{E}_{\bm{k},\eta'}\right)^2}{\left(\omega-\sigma_{3,\eta'}\mathcal{E}_{\bm{k},\eta'} + i \sigma_{3,\eta'}{\Gamma}_{\bm{k},\eta'}(\omega)\right)^2}
    \right]
    \right\}
    \tilde{c}_2(g(\beta\omega)),
\end{align}
and
\begin{align}    
    \label{appeq:Lxy-intra}
    L_{\lambda\lambda'}^{a;\mathrm{intra}} &= 
    - \frac{1}{\beta^{2}\hbar V} \sum_{\lambda''} \varepsilon_{\lambda\lambda'\lambda''}
    \sum_{\eta=1}^{N}\sum_{\bm{k}} \Omega_{\bm{k},\eta}^{\lambda''}
    \mathrm{P}\int_{-\infty}^{\infty} d\omega
    \frac{\varepsilon_{\bm{k},\eta}^3}{\omega^3}
    \rho_{\bm{k},\eta}(\omega) \tilde{c}_2(g(\beta\omega)),
\end{align}
respectively.

As discussed in Sec.~\ref{sec:appro-green}, we assume that the damping rate of magnons is small enough in the vicinity of the zero energy. 
This implies that the contribution at $\omega\simeq \varepsilon_{\bm{k},\eta}$, corresponding to the peak of $\rho_{\bm{k},\eta}(\omega)$, are dominant in the $\omega$ integrals of Eqs.~\eqref{appeq:Lxy-inter} and \eqref{appeq:Lxy-intra}.
Thus, we approximate $\omega$ appearing explicitly in these equations to $\omega\simeq \varepsilon_{\bm{k},\eta}$.
Furthermore, we incorporate contributions up to the first order of $\Gamma_{\bm{k},\eta}(\omega)$ into the thermal Hall conductivity.
Since the spectral function given in Eq.~\eqref{eq:spec-Green} does not contain zeroth order contributions of $\Gamma_{\bm{k},\eta}(\omega)$, we apply the following approximation $[\varepsilon_{\bm{k},\eta} - \sigma_{3,\eta'}\mathcal{E}_{\bm{k},\eta'} + i\sigma_{3,\eta'}\Gamma_{\bm{k},\eta'}]^{-x} \simeq [\varepsilon_{\bm{k},\eta} - \sigma_{3,\eta'}\mathcal{E}_{\bm{k},\eta'}]^{-x}$ for $x=1,2$ in Eqs.~\eqref{appeq:Lxy-inter} and \eqref{appeq:Lxy-intra}.
Using the above approximations, we find $L_{\lambda\lambda'}^{a;\mathrm{inter}}\simeq 0$ and the thermal Hall conductivity is represented as
\begin{align}
    \kappa_{\lambda\lambda'}^{H}
    \simeq \frac{L_{\lambda\lambda'}^{a;\mathrm{intra}}}{T}
    \simeq&
    -\frac{k_{\rm B}^2 T}{\hbar V} \sum_{\lambda''}\sum_{\eta=1}^{N}\sum_{\bm{k}}
    \varepsilon_{\lambda\lambda'\lambda''} \Omega_{\bm{k},\eta}^{\lambda''} \int_{-\infty}^{\infty} d\omega \rho_{\bm{k},\eta}(\omega) 
    \left[ c_{2}(g(\beta\omega)) - \frac{\pi^2}{3} \right].
\end{align}
Using the sum rule for the spectral function given in Eq.~\eqref{eq:sumrule-rho} and 
$\sum_{\eta=1}^{N}\sum_{\bm{k}}\Omega_{\bm{k},\eta}^{\lambda} = 0$, we find
\begin{align}
    \sum_{\eta=1}^{N} \sum_{\bm{k}} \Omega_{\bm{k},\eta}^{\lambda''} \int_{-\infty}^{\infty} \rho_{\bm{k},\eta} (\omega) 
    = 
    \sum_{\eta=1}^{N} \sum_{\bm{k}} \Omega_{\bm{k},\eta}^{\lambda''}
    = 0.
\end{align}
Therefore, $\kappa_{\lambda\lambda'}^{H}$ is given by
\begin{align}
    \kappa_{\lambda\lambda'}^{H}
    \simeq
    -\frac{k_{\rm B}^2 T}{\hbar V} \sum_{\lambda''}\sum_{\eta=1}^{N}\sum_{\bm{k}}
    \varepsilon_{\lambda\lambda'\lambda''} \Omega_{\bm{k},\eta}^{\lambda''} \int_{-\infty}^{\infty} d\omega \rho_{\bm{k},\eta}(\omega) 
    c_{2}(g(\beta\omega)).
\end{align}

\end{widetext}
\bibliography{./refs}
\end{document}